\documentclass[12pt]{article}
\usepackage[cp1250]{inputenc}
\usepackage[verbose,a4paper,tmargin=0.5in,lmargin=1in,rmargin=1in]{geometry}
\usepackage[pdftex]{graphicx}
\usepackage{amsfonts}
\usepackage{indentfirst}
\usepackage{latexsym}
\usepackage{amsmath}
\usepackage{amssymb}
\usepackage{psfrag}
\usepackage{eufrak}
\usepackage[mathscr]{eucal} 

\title{Homothetic Killing Vectors in Expanding $\mathcal{HH}$-Spaces with $\Lambda$.}

\author{$\textrm{Adam Chudecki}^{*} \dag$}

\begin{document}

\maketitle

$*$ Center of Mathematics and Physics  

$\ \ $ Technical University of Łódź           

$\ \ $ Al. Politechniki 11, 90-924 Łódź, Poland
\newline

$\dag$ E-mail: adam.chudecki@p.lodz.pl
\newline
\newline
\textbf{Abstract.} Conformal Killing equations and their integrability conditions for expanding hyperheavenly spaces with $\Lambda$ in spinorial formalism are studied. It is shown that any conformal Killing vector reduces to homothetic or isometric Killing vector. Reduction of respective Killing equation to one \textsl{master equation} is presented. Classification of homothetic and isometric Killing vectors is given. Type $\textrm{[D]} \otimes \textrm{[any]}$ is analysed in details and some expanding $\mathcal{HH}$ complex metrics of types $\textrm{[III, N]} \otimes \textrm{[III, N]}$ with $\Lambda$ admitting isometric Killing vectors are found.
\newline
\newline
\textbf{PACS numbers:} 04.20.Cv, 04.20.Jb, 04.20.Gz

\section{Introduction}

This paper is the second part of more extensive work devoted to the conformal, homothetic and isometric Killing symmetries in hyperheavenly spaces. In the previous paper \cite{biblio_51} the nonexpanding hyperheavenly spaces have been considered and it was a generalization of the \cite{biblio_27}. Now we deal with expanding case. 

The hyperheavenly spaces are the generalization of the heavenly spaces and they were discovered by Plebański and Robinson in 1976 \cite{biblio_18} - \cite{biblio_19}. The structure of the heavenly and hyperheavenly spaces, especially in the spinorial formalism was described in \cite{biblio_11} - \cite{biblio_26}.

The problem of Killing symmetries given by the set of equations $\nabla_{(a} K_{b)} = \chi \, g_{ab}$ in expanding hyperheavenly spaces was presented in \cite{biblio_52} and it is the unique paper devoted to this problem. However, in \cite{biblio_52} the authors considered only homothetic and isometric symmetries. Indeed, the conformal symmetries are not allowed by the nonexpanding hyperheavenly spaces but in \cite{biblio_52} it was set \textsl{a priori}. Here we prove that $\chi$ must be constant, what follows from the integrability conditions of the equations $\nabla_{(a} K_{b)} = \chi \, g_{ab}$. Moreover, considerations presented in \cite{biblio_52} does not include the cosmological constant. Our work fill that gap.

The way of reduction of the Killing problem in \cite{biblio_52} has been done in quite different way. We present the alternative way to obtain the master equation and its integrability conditions. It is the main reason, why the classification of the Killing vectors in our work differs of that given in \cite{biblio_52}. 

In the present paper we use the following terminology. Eq. (\ref{konforemne_rownanie_Killinga}) is called the conformal Killing equation and its solution $K$ the conformal Killing vector (as it was mentioned earlier, this case is not allowed in the expanding hyperheavenly spaces). If $\chi = \chi_{0} = \textrm{const}$ then Eq. (\ref{konforemne_rownanie_Killinga}) is called the homothetic Killing equation and $K$, the homothetic Killing vector. Finally, if $\chi=0$ we have the isometric Killing equation and the isometric Killing vector, respectively.

Note that the results of our paper can be quickly carry over to the case of real spacetimes of neutral signature $(++--)$. Such spaces have attracted a great deal of interest in Osserman space theory \cite{biblio_60, biblio_32, biblio_33}.

Our paper is organized as follows. In section 2 the general structure of expanding hyperheavenly space is presented. We recall the form of the hyperheavenly equation, connection forms and curvature. In section 3 the explicit form of the Killing equations and their integrability conditions are found and the general results, especially the form of the master equation and the transformation formulas are presented. Section 4 is devoted to the classification of the isometric and homothetic Killing vectors. In section 5 we find the reduced hyperheavenly equations for the type $\textrm{[D]} \otimes \textrm{[any]}$ admitting different isometric and homothetic Killing vectors. Some interesting metrics of the type $\textrm{[III, N]} \otimes \textrm{[III, N]}$ with nonzero cosmological constant are explicitly given. Some of them are examples of the Osserman spaces with the symmetry. Detailed reduction of the Killing problem for the expanding hyperheavenly spaces are done in section 6. Concluding remarks end our paper. 

Our work is motivated by still actual question: how one can generate new real Lorentzian metrics from holomorphic ones (the Plebański program). We guess that this is an important problem of mathematical relativity and that its solution will emerge from deeper and deeper insight into complex relativity. We suppose that the present paper is a step in this direction.

\section{Hyperheavenly spaces.}

\subsection{General structure of hyperheavenly spaces with $\Lambda$.}

\textsl{$\mathcal{HH}$-space with cosmological constant} is a 4 - dimensional complex analytic differential manifold $\mathcal{M}$ endowed with a holomorphic Riemannian metric $ds^{2}$ satisfying the vacuum Einstein equations with cosmological constant $\Lambda$ and such that the self - dual or anti - self - dual part of the Weyl tensor is algebraically degenerate \cite{biblio_18, biblio_19, biblio_20}. These kind of spaces admits a congruence of totally null, self-dual (or anti-self-dual, respectively) surfaces, called \textsl{null strings} \cite{biblio_28}. In this paper we deal with self-dual null strings. Coordinate system can be always chosen such that $\Gamma_{422}=\Gamma_{424}=0$ \cite{biblio_19}, the surface element of null string is given by $e^{1} \wedge e^{3}$, and null tetrad $(e^{1},e^{2},e^{3},e^{4})$ in spinorial notation can be chosen as
\begin{eqnarray}
\label{tetrada_spinorowa}
e_{\dot{A}} := \phi^{-2} \, dq_{\dot{A}} &=& 
\left[ \begin{array}{c}  e^{3}  \\ e^{1}  \end{array} \right]  \ \ = \ \ 
-\frac{1}{\sqrt{2}} \, g^{2}_{\ \, \dot{A}}
\\ \nonumber
E^{\dot{A}} := -dp^{\dot{A}} + Q^{\dot{A} \dot{B}} \, dq_{\dot{B}} &=&
\left[ \begin{array}{c}  e^{4}  \\ e^{2}  \end{array} \right]  \ \ = \ \ 
\frac{1}{\sqrt{2}} \, g^{1 \dot{A}}
\end{eqnarray}
where
\begin{equation}
(g^{A\dot{B}}) := \sqrt{2}
\left[\begin{array}{cc}
e^4 & e^2 \\
e^1 & -e^3
\end{array}\right] 
\end{equation}
$\phi$ and $Q^{\dot{A} \dot{B}}$ are holomorphic functions. Spinorial coordinates $p^{\dot{A}}$ are coordinates on null strings given by $q_{\dot{A}}=\textrm{const}$.
Define the following operators
\begin{eqnarray}
&&\partial_{\dot{A}} := \frac{\partial}{\partial p^{\dot{A}}}
\ \ \ \ \ \ \ \ \ \ \ \ \ \ \ \ \ \ \ \ \ \ \ \ 
\partial^{\dot{A}} := \frac{\partial}{\partial p_{\dot{A}}}
\\ \nonumber
&&\eth^{\dot{A}} := \phi^{2} \left( \frac{\partial}{\partial q_{\dot{A}}} + Q^{\dot{A} \dot{B}} \partial_{\dot{B}} \right)
\ \ \ \ \ \ \ \ 
\eth_{\dot{A}} := \phi^{2} \left( \frac{\partial}{\partial q^{\dot{A}}} - Q_{\dot{A}}^{\ \ \dot{B}} \partial_{\dot{B}} \right)
\end{eqnarray}
They constitute the basis dual to $(e_{\dot{A}}, E^{\dot{B}})$
\begin{equation}
\label{baza_dualna}
-\partial_{\dot{A}} = 
\left[ \begin{array}{c}  \partial_{4}  \\ \partial_{2}  \end{array} \right] 
 \ \ \ \ \ \ \ \ 
\eth^{\dot{A}} = 
\left[ \begin{array}{c}  \partial_{3}  \\ \partial_{1}  \end{array} \right] 
\end{equation}
Spinorial indices are to be manipulated according to the rules $q_{\dot{A}} = \ \in_{\dot{A} \dot{B}} q^{\dot{B}}$, $q^{\dot{A}} = q_{\dot{B}} \in^{\dot{B} \dot{A}}$. Then, the rules to raise and lower spinor indices in the case of objects from tangent space read $\partial^{\dot{A}} = \partial_{\dot{B}} \in^{\dot{A} \dot{B}}$, $\partial_{\dot{A}} = \in_{\dot{B} \dot{A}} \partial^{\dot{B}}$, $\eth^{\dot{A}} = \eth_{\dot{B}} \in^{\dot{A} \dot{B}}$, $\eth_{\dot{A}} = \in_{\dot{B} \dot{A}} \eth^{\dot{B}}$. As usual, $\in_{AB}$ and $\in_{\dot{A}\dot{B}}$ are spinorial Levi-Civita symbols
\begin{eqnarray}
&&( \in_{AB} ) := \left[ \begin{array}{cc}
                            0 & 1   \\
                           -1 & 0  
                            \end{array} \right] =: ( \in^{AB} )
\ \ \ , \ \ \ 
( \in_{\dot{A}\dot{B}} ) := \left[ \begin{array}{cc}
                            0 & 1   \\
                           -1 & 0  
                            \end{array} \right] =: ( \in^{\dot{A}\dot{B}} ) 
\\ \nonumber
&& \in_{AC} \in^{AB} = \delta^{B}_{C} \ \ \ , \ \ \ \in_{\dot{A}\dot{C}} \in^{\dot{A}\dot{B}} = \delta^{\dot{B}}_{\dot{C}} \ \ \ , \ \ \ 
(\delta^{A}_{C})= (\delta^{\dot{B}}_{\dot{C}})= \left[ \begin{array}{cc}
                            1 & 0   \\
                            0 & 1  
                            \end{array} \right]
\end{eqnarray}
The metric is given by
\begin{equation}
\label{metryka_ogolna}
ds^{2} =2 \, e^{1} \underset{s}{\otimes} e^{2} + 2 \, e^{3} \underset{s}{\otimes} e^{4}
= -\frac{1}{2} \, g_{A\dot{B}} \underset{s}{\otimes} g^{A\dot{B}}
= 2\phi^{-2} \, (- dp^{\dot{A}} \underset{s}{\otimes} dq_{\dot{A}} + 
           Q^{\dot{A} \dot{B}} \, dq_{\dot{A}} \underset{s}{\otimes} dq_{\dot{B}})
\end{equation}
The expansion 1-form which characterises congruence of self-dual null strings is defined by \cite{biblio_20}
\begin{equation}
\label{forma_ekspansji}
\theta := \theta_{a} e^{a} = \phi^{4} (\Gamma_{423} e^{3} + \Gamma_{421} e^{1}) = \phi \, \frac{\partial \phi}{\partial p_{\dot{A}}} \, dq_{\dot{A}}
\end{equation}
Consequently, we can consider two cases
\begin{itemize}
\item $\frac{\partial \phi}{\partial p_{\dot{A}}}=0 \rightarrow \theta=0$ and such a space is called \textsl{nonexpanding $\mathcal{HH}$-space} (in this case one can put $\phi=1$)
\item $\frac{\partial \phi}{\partial p_{\dot{A}}}\ne0 \rightarrow \theta \ne 0$; such a space is called  \textsl{expanding $\mathcal{HH}$-space}
\end{itemize}
In the present paper we deal with expanding case.
\newline
Important remark: In geometrical terms $\theta=0$ means that the null strings are parallely propagated and $\theta \ne 0$ means that they are not parallely propagated. Therefore, in the case of real spacetime of signature $(++--)$ the case $\theta=0$ corresponds to the case when $(\mathcal{M}, ds^{2})$ is a Walker space \cite{biblio_32}.

\subsection{Expanding hyperheavenly spaces.}

By reducing Einstein equations \cite{biblio_19, biblio_20} one gets
\begin{equation}
\phi=J_{\dot{A}}p^{\dot{A}}  
\end{equation}
where $J_{\dot{A}}$ is constant, nonzero spinor. Let $K_{\dot{A}}$ be a spinor, defined by the relation
\begin{equation}
K^{\dot{A}} J_{\dot{B}} - K_{\dot{B}} J^{\dot{A}} = \tau \, \delta^{\dot{A}}_{\dot{B}}
\ \ \ \ \textrm{where} \ \ \ \ \tau = K^{\dot{A}} J_{\dot{A}} \ne 0
\end{equation}
Then
\begin{subequations}
\begin{eqnarray}
\label{expanding_QAB}
Q^{\dot{A} \dot{B}} &=& -2 \, J^{(\dot{A}} \partial^{\dot{B})} W - \phi \, \partial^{\dot{A}} \partial^{\dot{B}} W+ \frac{\mu}{\tau^{2}} \phi^{3} K^{\dot{A}} K^{\dot{B}}+ \frac{\Lambda}{6\tau^{2}} K^{\dot{A}} K^{\dot{B}}
\\ 
\label{expanding_QAB_z_Omega}
&=& \partial^{(\dot{A}} \Omega^{\dot{B})} + \frac{\mu}{\tau^{2}} \phi^{3} K^{\dot{A}} K^{\dot{B}}
\end{eqnarray}
\end{subequations}
where
\begin{equation}
\label{postac_Omegi}
\Omega^{\dot{B}} 
= -\phi \, \partial^{\dot{B}} W - 3 J^{\dot{B}} \, W + \frac{\Lambda}{6\tau^{2}} K^{\dot{B}} \eta
= - \phi^{4} \, \partial^{\dot{B}} (\phi^{-3}W) + \frac{\Lambda}{6\tau^{2}} K^{\dot{B}} \eta
\end{equation}
$\mu=\mu(q^{\dot{N}})$ is an arbitrary function, $\Lambda$ is the cosmological constant and
\begin{equation}
\eta := K^{\dot{A}} p_{\dot{A}} \ \ \ \ \ \rightarrow \ \ \ \ \ \tau  p^{\dot{A}} = \eta  \, J^{\dot{A}} + \phi \, K^{\dot{A}} 
\end{equation}
Einstein equations can be reduced to one equation called the \textsl{expanding hyperheavenly equation with $\Lambda$}
\begin{eqnarray}
\label{expanding_hyperheavenly_equation_form1}
&&\mathcal{T} +(\phi^{-1}J^{\dot{A}} \, W_{p^{\dot{A}}})^{2} + \phi^{-1} W_{p_{\dot{A}}q^{\dot{A}}}
+ \mu \phi^{4} \, \partial_{\phi} (\phi^{-3}W) 
\\ \nonumber
&&+ \frac{\eta}{2 \tau^{2}} \big( \eta J^{\dot{C}} - \phi K^{\dot{C}} \big) \frac{\partial \mu}{\partial q^{\dot{C}}}
-\frac{\Lambda}{3} \, \phi^{-2} \, \partial_{\phi} W = N_{\dot{A}} \, p^{\dot{A}} + \gamma
\end{eqnarray}
\begin{equation}
\mathcal{T} := \frac{1}{2}\phi^{-2} Q^{\dot{A} \dot{B}} Q_{\dot{A} \dot{B}} 
\end{equation}
\begin{equation}
\partial_{\phi} = \frac{1}{\tau} K^{\dot{A}} \partial_{\dot{A}} 
\ \ \ \ \ \ \ \ \partial_{\eta} = \frac{1}{\tau} J^{\dot{A}} \partial_{\dot{A}} 
\end{equation}
The form (\ref{expanding_hyperheavenly_equation_form1}) will be especially useful in calculations in section \ref{sekcja_szczegolowe_obliczenia}. Inserting the explicit form of $\mathcal{T}$ into Eq. (\ref{expanding_hyperheavenly_equation_form1}), the expanding hyperheavenly equation with $\Lambda$ can be presented in the well know form \cite{biblio_21}
\begin{eqnarray}
\label{expanding_hyperheavenly_equation_form2}
&&\frac{1}{2} \phi^{4} (\phi^{-2}W_{p_{\dot{B}}})_{p_{\dot{A}}}
                     (\phi^{-2}W_{p^{\dot{B}}})_{p^{\dot{A}}}
+\phi^{-1} W_{p_{\dot{A}} q^{\dot{A}}} 
- \mu \phi^{4} \, [ \phi^{-1} (\phi^{-1}W)_{\phi} ]_{\phi}
\\ \nonumber
&&+ \frac{\eta}{2 \tau^{2}} \Big( \eta J^{\dot{C}} - \phi K^{\dot{C}} \Big) \mu_{q^{\dot{C}}} - \frac{\Lambda}{6} \, \phi^{-1} \, W_{\phi \phi}
= N_{\dot{A}} \, p^{\dot{A}} + \gamma
\end{eqnarray}
where $N^{\dot{A}}$ and $\gamma$ are arbitrary functions of $q^{\dot{C}}$ only (constant on each null string), $W$ is \textsl{the key function}, and $W_{p^{\dot{A}}} \equiv \frac{\partial W}{\partial p^{\dot{A}}}$,  $W_{q_{\dot{A}}} \equiv \frac{\partial W}{\partial q_{\dot{A}}}, etc.$.
Explicit formulas for the connection 1-forms in spinorial formalism read
\begin{eqnarray}
\label{expanding_koneksja}
\mathbf{\Gamma}_{11}  &=& -\phi^{-1} J_{\dot{A}} \, e^{\dot{A}}                    
\\ \nonumber
\mathbf{\Gamma}_{12} &=& 
-\frac{1}{2} \phi \, \partial^{\dot{A}} (\phi \, Q_{\dot{A} \dot{B}}) \, e^{\dot{B}}
- \frac{3}{2} \phi^{-1} J_{\dot{B}} \, E^{\dot{B}}
\\ \nonumber
\mathbf{\Gamma}_{22} &=& -\phi^{2} \, (\eth^{\dot{A}} Q_{\dot{A}\dot{B}}) \, e^{\dot{B}} 
- \phi \, (Q_{\dot{A}\dot{B}}J^{\dot{A}}) E^{\dot{B}}
\\ \nonumber
\mathbf{\Gamma}_{\dot{A} \dot{B}} &=& 
\phi \Big( \phi \, \partial_{(\dot{A}} Q_{\dot{B})\dot{C}} + J^{\dot{D}} \in_{\dot{C} ( \dot{B}} Q_{\dot{A} ) \dot{D}} \Big)  e^{\dot{C}} + \phi^{-1} J_{(\dot{A}} E_{\dot{B})}
\end{eqnarray}
where
\begin{equation}
Q_{\dot{A}\dot{B}} J^{\dot{A}} =- \phi^{2} \, \left( J^{\dot{A}} (\phi^{-1} W_{p^{\dot{A}}})_{p^{\dot{B}}} + \frac{\mu}{\tau} \phi K_{\dot{B}} + \frac{\Lambda}{6 \tau} \phi^{-2} K_{\dot{B}}  \right)
= \phi^{-2} \, \eth_{\dot{B}} \phi
\end{equation}
and (as a consequence of hyperheavenly equation, especially useful in many calculations)
\begin{equation}
\eth^{\dot{A}} Q_{\dot{A}\dot{B}} = \phi^{4} \bigg\{ N_{\dot{B}} + J^{\dot{A}} \phi^{-2} W_{q^{\dot{A}}p^{\dot{B}}} + \left( \frac{1}{2 \tau} K^{\dot{C}} \in_{\dot{B} \dot{A}} - \frac{1}{\tau^{2}} K_{\dot{B}} K_{\dot{A}} J^{\dot{C}} \right) p^{\dot{A}} \frac{\partial \mu}{\partial q^{\dot{C}}} \bigg\}
\end{equation}
Decomposing the connection 1-forms according to
\begin{equation}
\label{rozklad_koneksji}
\mathbf{\Gamma}_{AB} = -\frac{1}{2} \, \mathbf{\Gamma}_{AB C \dot{D}} \, g^{C \dot{D}}
\ \ \ \ \ \ \ \ \  
\mathbf{\Gamma}_{\dot{A} \dot{B}} = -\frac{1}{2} \, \mathbf{\Gamma}_{\dot{A} \dot{B} C \dot{D}} \, g^{C \dot{D}}
\end{equation}
we get
\begin{eqnarray}
&&\mathbf{\Gamma}_{111 \dot{D}} =0  \ \ \ \ \ \ \ \ \ \ \ \ \ \ \ \ \  \ \ \ \ \ \ \ \ \ \ \ \ \ 
\mathbf{\Gamma}_{112 \dot{D}} = -\sqrt{2} \, \phi^{-1} \, J_{\dot{D}}
\\ \nonumber
&&\mathbf{\Gamma}_{121 \dot{D}} = \frac{3}{\sqrt{2}} \phi^{-1} \, J_{\dot{D}} \ \ \ \ \ \ \ \ \ \ \ \ \ \ \ \ \ \ 
  \mathbf{\Gamma}_{122 \dot{D}} = - \frac{1}{\sqrt{2}} \, \phi \, \partial^{\dot{A}} (\phi \, Q_{\dot{A}\dot{D}})
\\ \nonumber
&& \mathbf{\Gamma}_{221 \dot{D}} = \sqrt{2} \phi \, Q_{\dot{D}\dot{A}} J^{\dot{A}}
\ \ \ \ \ \ \ \ \ \ \ \ \ \ \ \,
\mathbf{\Gamma}_{222 \dot{D}} = -\sqrt{2} \phi^{2} \, \eth^{\dot{A}} Q_{\dot{A}\dot{D}}
\\ \nonumber
&&\mathbf{\Gamma}_{\dot{A} \dot{B} 1 \dot{D}} = -\sqrt{2} \phi^{-1} \, J_{(\dot{A}} \in_{\dot{B})\dot{D}}
\ \ \ \ \ \ 
\mathbf{\Gamma}_{\dot{A} \dot{B} 2 \dot{D}} =  \sqrt{2} \, \phi \Big(  
\phi \, \partial_{(\dot{A}} Q_{\dot{B})\dot{D}} +  \in_{\dot{D}(\dot{B}} Q_{\dot{A})\dot{C}} J^{\dot{C}} \Big) 
\end{eqnarray}
The conformal curvature is given by
\begin{eqnarray}
\label{expanding_krzywizna}
C^{(5)}  &=& 0
\\ \nonumber
C^{(4)}  &=& 0
\\ \nonumber
C^{(3)}  &=& - 2 \mu \, \phi^3 
\\ \nonumber
C^{(2)}  &=&  2\phi^{5} \left(N_{\dot{A}}J^{\dot{A}} -  p^{\dot{A}}  \frac{\partial \mu}{\partial q^{\dot{A}}} \right) = 2\phi^{5} \left(-\frac{1}{2} \nu \tau - p^{\dot{A}}   \mu_{q^{\dot{A}}} \right)
\\ \nonumber
C^{(1)} &=& 2 \phi^{7}  \Bigg\{
\frac{1}{2} \mu \varkappa \, \phi^{3} 
- \frac{\tau}{2} \, \nu_{q_{\dot{A}}}   p_{\dot{A}}
+\frac{\mu}{2 \tau^{2}} \eta \phi^{3} K^{\dot{R}} \mu_{q^{\dot{R}}}
- \frac{\Lambda}{6 \tau^{2}} \bigg( \tau^{2} \varkappa + \eta \, K^{\dot{R}} 
\mu_{q^{\dot{R}}}  \bigg)
\\ \nonumber
&& \ \ \ \ \ \  - \frac{1}{2} \, \mu_{q^{\dot{C}} q^{\dot{D}}} p^{\dot{D}}  p^{\dot{C}} 
+J^{\dot{B}} \frac{\partial}{\partial q^{\dot{B}}} 
\left(  \gamma +3 \mu \, W \right) 
\\ \nonumber
&&  \ \ \ \ \ \ + \bigg(
-\tau \nu \,  J_{\dot{B}} - \left( p_{\dot{B}} J^{\dot{C}} + J_{\dot{B}} p^{\dot{C}}  \right) \mu_{q^{\dot{C}}} 
\bigg)  W_{p_{\dot{B}}}
\Bigg\}
\\ \nonumber
C_{\dot{A}\dot{B}\dot{C}\dot{D}} &= &
\phi^{3} 
\left( W-\frac{\mu}{4 \tau^2} \, \eta^2 \phi^2     \right)_{p^{\dot{A}}p^{\dot{B}}p^{\dot{C}}p^{\dot{D}}}
\end{eqnarray}
where we decompose $N_{\dot{A}}$ according to the formula
\begin{equation}
2N_{\dot{A}} =: \nu K_{\dot{A}} + \varkappa J_{\dot{A}}
\end{equation}
which defines the quantities $\nu=\nu(q_{\dot{B}})$ and $\varkappa=\varkappa(q_{\dot{B}})$
\begin{equation}
\nu :=  -\frac{2}{\tau}   N_{\dot{A}} J^{\dot{A}}
\ \ \ \ \ \ \ \ 
\varkappa :=  \frac{2}{\tau}   N_{\dot{A}} K^{\dot{A}}
\end{equation}
After substituting (\ref{expanding_QAB}) into (\ref{metryka_ogolna}) we obtain
\begin{eqnarray}
\label{metrykaprzestrzeni_HH_z_ekspanja}
ds^{2} &=& (\phi \tau)^{-2} \bigg\{ 2\tau (d \eta \underset{s}{\otimes} d w - d \phi \underset{s}{\otimes} dt) + 2 \, \Big( -\tau^{2} \phi \, W_{\eta\eta} + \mu \phi^{3}+\frac{\Lambda}{6}  \Big) 
dt \underset{s}{\otimes}dt \ \ \ \ \ \ 
\\ \nonumber
&& \ \ \ \ \ \ \ \ \ \ +4 \left( -\tau^{2} \phi \, W_{\eta \phi} + \tau^{2} \, W_{\eta}\right) dw \underset{s}{\otimes}dt
  + 2 \left(  -\tau^{2} \phi \, W_{\phi \phi} + 2\tau^{2} \, W_{\phi}         \right) dw \underset{s}{\otimes}dw \bigg\} \ \ \ \ \ \ 
\end{eqnarray}
where
\begin{equation}
t := K^{\dot{A}}q_{\dot{A}}
\ \ \ \ \ \ \ \ \ \ \ \
w := J_{\dot{A}}q^{\dot{A}}
\ \ \ \ \ \rightarrow \ \ \ \ \ 
q_{\dot{B}} = \frac{1}{\tau} \, (t J_{\dot{B}} + w K_{\dot{B}})
\end{equation}
\label{rownanie_HH_we_wspolrzednych}
In $(\phi, \eta, w,t)$-language, the hyperheavenly equation reads
\begin{eqnarray}
&& \tau^{2} \Big( W_{\eta \eta}W_{\phi \phi} - W_{\eta \phi}W_{\eta \phi} + 2 \phi^{-1} W_{\eta} W_{\eta \phi} - 2 \phi^{-1} W_{\phi}W_{\eta\eta} \Big) + \tau \phi^{-1} \Big( W_{w\eta}-W_{t\phi} \Big)
\ \ \ \ \ \ 
\\ \nonumber
&&- \mu \Big( \phi^{2} W_{\phi\phi} - 3 \phi W_{\phi} + 3 W \Big) + \frac{\eta}{2 \tau} (\mu_{t} \eta - \mu_{w} \phi) - \frac{\Lambda}{6} \phi^{-1} W_{\phi \phi} = \frac{1}{2} \varkappa \phi - \frac{1}{2} \nu \eta + \gamma
\end{eqnarray}
[Not loosing generality one can set $\tau=1$.]

\subsection{Gauge freedom.}
\label{subsection_gauge}

The problem of coordinate gauge freedom in expanding hyperheavenly spaces is similar to the one in nonexpanding spaces (compare with \cite{biblio_51, biblio_19, biblio_20} and in the final form with $\Lambda$ with \cite{biblio_33}). The form of metric (\ref{metryka_ogolna}) admits the coordinate gauge freedom
\begin{eqnarray}
\label{gaug_wspolrzednych}
q'_{\dot{A}} &=& q'_{\dot{A}} (q_{\dot{B}})
\\ \nonumber
p'^{\dot{A}} &=& \lambda^{-1} \, \frac{\partial q_{\dot{B}}}{\partial q'_{\dot{A}}} \, p^{\dot{B}} + \sigma^{\dot{A}} 
\ \ \ \ , \ \ \ \ 
\lambda=\lambda(q_{\dot{A}}), \ \ \sigma^{\dot{A}}=\sigma^{\dot{A}} (q_{\dot{B}})
\end{eqnarray}
where $\lambda$ and $\sigma^{\dot{A}}$ are arbitrary functions of $q^{\dot{A}}$.
\newline
The functions $\phi$ and $Q^{\dot{A}\dot{B}}$ transform under (\ref{gaug_wspolrzednych}) as follows
\begin{equation}
\phi' = \lambda^{-\frac{1}{2}} \, \phi \ \ \ \ , \ \ \ \ 
Q'^{\dot{A}\dot{B}} = \lambda^{-1} \, \frac{\partial q_{\dot{C}}}{\partial q'_{\dot{A}}} \,
     \frac{\partial q_{\dot{D}}}{\partial q'_{\dot{B}}} \, Q^{\dot{C}\dot{D}} + 
   \frac{\partial q_{\dot{C}}}{\partial q'_{(\dot{A}}} \, 
    \frac{\partial p'^{\dot{B})}}{\partial q_{\dot{C}}} 
\end{equation}
Denote 
\begin{eqnarray}
D_{\dot{A}}^{\ \ \dot{B}} 
&=& \frac{\partial q'_{\dot{A}}}{\partial q_{\dot{B}}} 
= \Delta \, \frac{\partial q^{\dot{B}}}{\partial q'^{\dot{A}}}
= \lambda \Delta \, \frac{\partial p'_{\dot{A}}}{\partial p_{\dot{B}}}
= \lambda^{-1} \frac{\partial p^{\dot{B}}}{\partial p'^{\dot{A}}}
\\ \nonumber
D^{-1 \ \ \dot{B}}_{\ \ \; \dot{A}} 
&=& \frac{\partial q_{\dot{A}}}{\partial q'_{\dot{B}}}
= \Delta^{-1} \, \frac{\partial q'^{\dot{B}}}{\partial q^{\dot{A}}}
= \lambda \, \frac{\partial p'^{\dot{B}}}{\partial p^{\dot{A}}}
= \lambda^{-1} \Delta^{-1} \, \frac{\partial p_{\dot{A}}}{\partial p'_{\dot{B}}}
\end{eqnarray}
where $\Delta$ is the determinant
\begin{equation}
\Delta := \det 
\left( \frac{\partial q'_{\dot{A}}}{\partial q_{\dot{B}}} \right)
 = \frac{1}{2} \, D_{\dot{A}\dot{B}} D^{\dot{A}\dot{B}}
\end{equation}
It is easy to see that the transformation (\ref{gaug_wspolrzednych}) is equivalent to the spinorial transformation of the tetrad with 
\begin{eqnarray}
\label{transformacja_spinorowa}
L^{A}_{\ \; B} &=& \left[ \begin{array}{cc}
       \lambda^{-1}\Delta^{-\frac{1}{2}} & h\phi^{2}  \Delta^{\frac{1}{2}}   \\
       0 &  \lambda \Delta^{\frac{1}{2}}  
       \end{array} \right]
\\ \nonumber
M^{\dot{A}}_{\ \; \dot{B}} &=& \Delta^{\frac{1}{2}} \, D^{-1 \ \ \dot{A}}_{\ \ \; \dot{B}}
\end{eqnarray}
where
\begin{equation}
2h := - \lambda^{-1} \, D^{-1 \ \ \dot{R}}_{\ \ \; \dot{S}} \, 
\frac{\partial p^{\dot{S}}}{\partial q'^{\dot{R}}} 
= \frac{1}{\Delta} \, D_{\dot{S}}^{\ \ \dot{R}} \, 
\frac{\partial p'^{\dot{S}}}{\partial q^{\dot{R}}} 
= \frac{\partial \sigma^{\dot{R}}}{\partial q'^{\dot{R}}} + \frac{1}{\Delta} \frac{\partial (\lambda^{-1})}{\partial q^{\dot{R}}} p^{\dot{R}}
\end{equation}
[If $\Psi^{A...\dot{B}...}$ is a spinorial quantity, it transforms according to formula 
\begin{equation}
\nonumber
\Psi'^{A...\dot{B}...}=L^{A}_{\ \; R}...M^{\dot{B}}_{\ \; \dot{S}}...\Psi^{R...\dot{S}...} \ \ \ \ ]
\end{equation}
\newline
Performing reduction of Einstein equations to the hyperheavenly equation it is convenient to maintain $J'^{\dot{A}} = J^{\dot{A}}=\textrm{const}$ and $K'^{\dot{A}} = K^{\dot{A}}=\textrm{const}$. It puts some restrictions on the gauge freedom. After some analysis one finds that the gauge freedom is restricted to the following transformations
\begin{eqnarray}
\label{jawna_postac_gaugu}
&&w'=w'(w) \ \ \ , \ \ \ t'=t'(w,t) \ \ \ , \ \ \ t'_{t} = \lambda^{-\frac{1}{2}} \ \ \ \rightarrow
\\ \nonumber
&&D_{\dot{A}}^{\ \ \dot{B}} = \frac{\partial q'_{\dot{A}}}{\partial q_{\dot{B}}} = - \frac{1}{\tau} \, (t'_{w} J_{\dot{A}} + w'_{w} K_{\dot{A}} ) J^{\dot{B}} 
+ \frac{1}{\tau} \, \lambda^{-\frac{1}{2}} J_{\dot{A}}  K^{\dot{B}}
\\ \nonumber
&&D^{-1 \ \ \dot{B}}_{\ \ \; \dot{A}} = \frac{1}{\tau} \, \bigg( \lambda^{\frac{1}{2}} \frac{t'_{w}}{w'_{w}} J_{\dot{A}} - \frac{1}{w'_{w}} K_{\dot{A}} \bigg) J^{\dot{B}} + \frac{1}{\tau} \lambda^{\frac{1}{2}} J_{\dot{A}}K^{\dot{B}}
\\ \nonumber
&&\sigma^{\dot{A}} = \sigma J^{\dot{A}}  \ \ \ \ \ \ , \ \ \ \ \ \ \Delta=w'_{w} \lambda^{-\frac{1}{2}}
\end{eqnarray}
where $\sigma=\sigma(w,t)$ is an arbitrary function and, of course,
\begin{equation}
\partial_{t} = \frac{1}{\tau} J^{\dot{A}} \frac{\partial}{\partial q^{\dot{A}}} \ \ \ \ \ , \ \ \ \ \ 
\partial_{w} = \frac{1}{\tau} K^{\dot{A}} \frac{\partial}{\partial q^{\dot{A}}}
\end{equation}
Straightforward but long calculations show that the function $W$ transforms under (\ref{jawna_postac_gaugu}) as follows \cite{biblio_33}
\begin{eqnarray}
(w'_{w})^{2} \lambda^{\frac{3}{2}} W' &=& W + \frac{\mu}{2 \tau^{2}} \lambda^{\frac{1}{2}} t'_{w} \Big(\phi^{3} \eta + \frac{1}{2} \lambda^{\frac{1}{2}} t'_{w} \phi^{4} \Big) - \frac{1}{3} L \, \phi^{3}
\\ \nonumber
&&- \frac{1}{2\tau} \lambda^{\frac{1}{2}} w'_{w} \, 
\frac{\partial}{\partial q_{(\dot{A}}} 
\bigg( \frac{1}{w'_{w}} \frac{\partial t'}{\partial q_{\dot{B})}} \bigg) \, p_{\dot{A}} p_{\dot{B}} 
\\ \nonumber
&&+ \bigg(  -\frac{1}{2} \lambda w'_{w} \, \frac{\partial \sigma}{\partial q_{\dot{A}}}
+\frac{\Lambda}{12 \tau^{2}} \, 
(\lambda {t'_{w}}^{2} \, J^{\dot{A}} - 2 \lambda^{\frac{1}{2}} t'_{w} \, K^{\dot{A}}) \bigg) \,  p_{\dot{A}} - M
\end{eqnarray}
where $M$ and $L$ are arbitrary functions of $q_{\dot{A}}$ only, giving an additional gauge freedom for $W$. Similarly, for the structural functions $\mu'$, $\nu'$, $\varkappa'$ and $\gamma'$ one gets \cite{biblio_33}
\begin{subequations}
\begin{eqnarray}
\label{trans_mu}
\lambda^{-\frac{3}{2}} \, \mu' &=&   \mu
\\
\label{trans_nu}
w'_{w} \lambda^{-1} \, \nu' &=& \nu -2\lambda w'_{w} \sigma \mu_{t} - 3 \mu w'_{w}  \, (\sigma \lambda)_{t}
\\
\label{trans_varkappa}
(w'_{w})^{2} \lambda^{-\frac{1}{2}} \varkappa' &=& \varkappa +\lambda^{\frac{1}{2}} t'_{w} \, \nu +2\tau \, L_{t} 
\\ \nonumber
&& +\lambda^{\frac{1}{2}} \sigma w'_{w} 
\Big( \lambda \, (t'_{w} \mu_{t} - \lambda^{-\frac{1}{2}} \mu_{w}) 
+\frac{3}{2} \, \mu \, (t'_{w} \lambda_{t} - \lambda^{-\frac{1}{2}} \lambda_{w} ) \Big)
\\
\label{trans_gamma}
(w'_{w})^{2} \gamma' &=& \gamma + 3 \mu M - (w'_{w})^{\frac{1}{2}} [(w'_{w})^{-\frac{1}{2}}]_{ww} + \frac{1}{2} \, \lambda w'_{w} \tau \nu \sigma
\\ \nonumber
&&-\frac{3}{2} \, (\lambda w'_{w})^{2} \mu \tau \sigma \sigma_{t}
- \frac{1}{4} \, \tau \lambda (w'_{w} \sigma)^{2} (3 \mu \lambda_{t} + 2 \lambda \mu_{t})
+ \frac{\Lambda}{3} L
\end{eqnarray}
\end{subequations}
Finally
\begin{equation}
\eta' = \frac{\lambda^{-1}}{w'_{w}} \eta + \lambda^{-\frac{1}{2}} \frac{t'_{w}}{w'_{w}} \phi + \tau \sigma
= \frac{\lambda^{-\frac{1}{2}}}{w'_{w}} \big( t'_{t} \, \eta +t'_{w} \, \phi \big)  + \tau \sigma
\end{equation}

\subsection{Simplifications for concrete algebraic types.}

It follows from (\ref{trans_varkappa}) that for all types the structural function $\varkappa$ can be always gauged away by appropriate choice of the gauge function $L$. 

The types $[\textrm{II, D}] \otimes [\textrm{any}]$ are characterised by $C^{(3)} \ne 0$, so the structural function $\mu$ must be nonzero. However, by using the $\lambda$ transformation it can be always gauged to constant value. With nonzero, constant $\mu$, the function $\nu$ can be gauged to zero (compare (\ref{trans_nu})). It corresponds to the choice of tetrad in which $C^{(2)}=0$. The last step is to remove $\gamma$ with help of $M$ (\ref{trans_gamma}). Summing up, in the cases $[\textrm{II, D}] \otimes [\textrm{any}]$, not losing generality but only fixing coordinate gauge freedom one can always put $\nu=\varkappa=\gamma=0$, and $\mu=\mu_{0}=\textrm{const}$. Coefficient $C^{(1)}=6 \mu_{0} \phi^{7} J^{\dot{A}} W_{q^{\dot{A}}}=6 \mu_{0} \phi^{7} \tau W_{t}$. The well known, type $[\textrm{D}] \otimes [\textrm{any}]$ condition, $0= 2C^{(2)}C^{(2)}-3C^{(1)}C^{(3)}$, yields $C^{(1)}=0$, so with these choice of tetrad, for the type $[\textrm{D}] \otimes [\textrm{any}]$, the key function $W$ is a function of three variables only, $W=W(\phi, \eta, w)$.

For the type $[\textrm{III}] \otimes [\textrm{any}]$, $C^{(3)}=0$, so $\mu=0$. Employing the formulae  (\ref{trans_varkappa}) and (\ref{trans_gamma}) one finds, that a gauge freedom can be used to set $\varkappa=\gamma=0$, and $\nu=\nu_{0} = \textrm{const}$.

For type $[\textrm{N}] \otimes [\textrm{any}]$, $C^{(3)}=C^{(2)}=0 \ \ \rightarrow \ \ \mu=\nu=0$. With $\varkappa=0$, $C^{(1)}= 2\phi^{7} \tau \gamma_{t}$, so the type $[\textrm{N}] \otimes [\textrm{any}]$ is characterised by $\gamma_{t} \ne 0$.

If additionally $C^{(1)}=0$ then $\gamma=\gamma(w)$ and it can be gauged away with help of $w'_{w}$ or $L$ (compare (\ref{trans_gamma})). Consequently (see (\ref{expanding_krzywizna})) $C_{ABCD}=0$, but when $\Lambda \ne 0$ there does not exist any nonexpanding congruence of self-dual null strings. Such spaces admit conformal Killing symmetries (as we will show in section \ref{sekcja_Killing}) and we are going to consider them in all details in the next paper.

We summarize the above considerations in the following table
\newline
\newline
\begin{tabular}{|c|c|c|c|}   \hline
$\textrm{Structural}$ & $[\textrm{II}] \otimes [\textrm{any}]$  
& $[\textrm{III}] \otimes [\textrm{any}]$ & $[\textrm{N}] \otimes [\textrm{any}]$   \\  
 $\textrm{functions}$ & $[\textrm{D}] \otimes [\textrm{any}]$  &  &  
\\ \hline     
$\mu$ & $\ \ \ \ \mu_{0} = \textrm{const} \ne 0 \ \ \ \ $ & $0$   & $0$ 
\\ \hline
$\nu$ & $0$ & $\ \ \ \ \ \ \, \ \nu_{0} = \textrm{const} \ne 0 \ \ \, \  \ \ \ \ $   & $0$  
\\ \hline
$\varkappa$ & $0$ & $0$   & $0$ 
\\ \hline
$\gamma$ & $0$ & $0$   & $ \  \gamma=\gamma(t,w) \; \ \ , \ \  \gamma_{t} \ne 0 $ 
\\ \hline
\end{tabular}
\newline
\newline
Remark: in section \ref{sekcja_klasyfikacja} we deal with the classification of Killing vectors and in some cases in order to simplify the forms of these vectors we choose the gauges different of these presented in the above table.

\setlength\arraycolsep{2pt}
\setcounter{equation}{0}

\section{Conformal Killing symmetries.}
\label{sekcja_Killing}

\subsection{Conformal Killing equations and their integrability conditions in spinorial formalism.}

Define the spin-tensor $g^{aA \dot{B}}$ by the relation $g^{A \dot{B}} = g_{a}^{\ A \dot{B}} \, e^{a}$. Hence, $-\frac{1}{2} g^{aA\dot{B}}g_{bA\dot{B}} = \delta^{a}_{b}$ and $-\frac{1}{2}g^{aA\dot{B}}g_{aC\dot{D}} = \delta^{A}_{C} \delta^{\dot{B}}_{\dot{D}}$. The operators $\partial^{A\dot{B}}$ and $\nabla^{A\dot{B}}$ are spinorial equivalences of operators $\partial^{a}$ and $\nabla^{a}$, respectively, given by
\begin{equation}
\partial^{A\dot{B}} = g_{a}^{\  A \dot{B}} \partial^{a}  \ \ \ \ \ \ \ \ \ \ \ \ \ \ 
\nabla^{A\dot{B}} = g_{a}^{\  A \dot{B}} \nabla^{a} 
\end{equation}
In the basis (\ref{baza_dualna})
\begin{equation}
\label{rozklad_operatora}
\partial^{A\dot{B}} = \sqrt{2} \, (\delta^{A}_{1} \, \eth^{\dot{B}} - \delta^{A}_{2} \, \partial^{\dot{B}}) = \sqrt{2} \, [ \eth^{\dot{B}}, -\partial^{\dot{B}}]
\end{equation}
Killing vector has the form
\begin{equation}
K = K^{a} \, \partial_{a} = -\frac{1}{2} K_{A\dot{B}} \partial^{A\dot{B}} = k_{\dot{B}} \eth^{\dot{B}} - h_{\dot{B}} \partial^{\dot{B}}
\end{equation}
where we use the decomposition
\begin{equation}
\label{rozklad_wektora_Killinga}
K_{A\dot{B}} = - \sqrt{2} \, (\delta^{1}_{A} \,k_{\dot{B}} + \delta^{2}_{A} \,h_{\dot{B}} ) \,  
= - \sqrt{2} \,  [k_{\dot{B}},h_{\dot{B}}]
\end{equation}
Components of the Killing vector, $K^{a}$ and $K_{A \dot{B}}$ are related by the condition
\begin{equation}
K^{a} = - \frac{1}{2} g^{a A \dot{B}} \, K_{A \dot{B}} \ \ \leftrightarrow \ \ 
K_{A \dot{B}} = g_{a  A \dot{B}} \, K^{a}
\end{equation}
Conformal Killing equations with a conformal factor $\chi$ read
\begin{equation}
\label{konforemne_rownanie_Killinga}
\nabla_{(a} K_{b)} = \chi \, g_{ab}
\end{equation}
In spinorial form
\begin{equation}
\nabla_{A}^{\ \; \dot{B}} K_{C}^{\ \dot{D}} + \nabla_{C}^{\ \; \dot{D}} K_{A}^{\ \dot{B}} = -4 \chi \in_{AC} \in^{\dot{B}\dot{D}}
\end{equation}
what is equivalent to the following equations
\begin{subequations}
\begin{eqnarray}
\label{rozklad_rownania_Killinga}
E_{AC}^{\ \ \ \dot{B}\dot{D}} &\equiv& \nabla_{(A}^{\ \; \dot{(B}} K_{C)}^{\ \; \dot{D)}} = 0 
\\
\label{definicja_chi}
E &\equiv& \nabla^{N\dot{N}} K_{N\dot{N}} +8 \chi = 0
\end{eqnarray}
\end{subequations}
From (\ref{rozklad_rownania_Killinga}) and (\ref{definicja_chi}) it follows that
\begin{equation}
\nabla_{A}^{\ \; \dot{B}} K_{C}^{\ \; \dot{D}} = l_{AC} \in^{\dot{B}\dot{D}} + l^{\dot{B}\dot{D}} \in_{AC} - 2 \chi  \in_{AC} \in^{\dot{B}\dot{D}}
\end{equation}
with
\begin{equation}
l_{AC} := \frac{1}{2} \nabla_{(A}^{\ \ \dot{N}} K_{C)\dot{N}} \ \ \ \ \ \ \ \ \ \ \ \ 
l^{\dot{B}\dot{D}} := \frac{1}{2} \nabla^{N(\dot{B}} K_{N}^{\ \; \dot{D})}
\end{equation}
In \cite{biblio_26} the integrability conditions of (\ref{rozklad_rownania_Killinga}) and (\ref{definicja_chi}) have been found. For the Einstein space ($C_{AB\dot{C}\dot{D}}=0$, $R=-4 \Lambda$) these conditions consist of the following equations
\begin{subequations}
\begin{eqnarray}
\label{integrability_c_L}
L_{RST}^{\ \ \ \ \; \dot{A}} &\equiv& \nabla_{R}^{\ \; \dot{A}} l_{ST} + 2C^{N}_{\ RST} K_{N}^{\ \; \dot{A}} + \frac{2}{3} \Lambda \in_{R(S}K_{T)}^{\ \ \dot{A}} + 2 \in_{R(S} \nabla_{T)}^{\ \ \dot{A}} \chi =0 \ \ \ \ \ \ \ \ 
\\
\label{integrability_c_M}
M_{ABCD} &\equiv& K_{N\dot{N}} \nabla^{N\dot{N}}C_{ABCD} + 4C^{N}_{\ \; (ABC} l_{D)N} - 4 \chi C_{ABCD} = 0
\\
\label{integrability_c_N}
N_{AB}^{\ \ \ \dot{A}\dot{B}} &\equiv& \nabla_{A}^{\ \; \dot{A}} \nabla_{B}^{\ \; \dot{B}} \chi - \frac{2}{3} \Lambda \chi \in_{AB} \in^{\dot{A}\dot{B}} =0
\\
\label{integrability_c_R}
R_{ABC}^{\ \ \ \ \ \dot{A}} &\equiv& C^{N}_{\ \; ABC} \nabla_{N}^{\ \; \dot{A}} \chi =0
\end{eqnarray}
\end{subequations}
for undotted $l_{AB}$ and $C_{ABCD}$ and the respective equations $L_{\dot{R}\dot{S}\dot{T}}^{\ \ \ \ \; A}$, $M_{\dot{A}\dot{B}\dot{C}\dot{D}}$ and $R_{\dot{A}\dot{B}\dot{C}}^{\ \ \ \ \ A}$ for dotted $l_{\dot{A}\dot{B}}$ and $C_{\dot{A}\dot{B}\dot{C}\dot{D}}$.

\subsection{Explicit form of the conformal Killing equations and their integrability conditions in expanding hyperheavenly spaces with $\Lambda$.}

Equations (\ref{rozklad_rownania_Killinga}) - (\ref{definicja_chi}) together with their integrability conditions (\ref{integrability_c_L}) - (\ref{integrability_c_R}) form our problem to be solved. Using the formula for the spinorial covariant derivative
\begin{eqnarray}
\nabla_{M \dot{N}} \Psi^{A \dot{B}}_{C \dot{D}} &=& \partial_{M \dot{N}} \Psi^{A \dot{B}}_{C \dot{D}} 
+ \mathbf{\Gamma}^{A}_{\ SM \dot{N}} \, \Psi^{S \dot{B}}_{C \dot{D}} 
- \mathbf{\Gamma}^{S}_{\ CM \dot{N}} \, \Psi^{A \dot{B}}_{S \dot{D}}
\\ \nonumber
&&+ \mathbf{\Gamma}^{\dot{B}}_{\ \dot{S}M \dot{N}} \, \Psi^{A \dot{S}}_{C \dot{D}}
- \mathbf{\Gamma}^{\dot{S}}_{\ \dot{D}M \dot{N}} \, \Psi^{A \dot{B}}_{C \dot{S}}
\end{eqnarray}
the decomposition (\ref{rozklad_koneksji}), the formulae (\ref{rozklad_operatora}) and (\ref{rozklad_wektora_Killinga}), after some work one obtains the system of equations: 
\newline
Conformal Killing equations
\begin{eqnarray}
\label{rowwnania_Killinga}
E_{11}^{\ \ \dot{A}\dot{B}} &\equiv& 2 \phi^{-2} \, \partial^{(\dot{A}} \big( \phi^{2} k^{\dot{B})} \big) = 0
\\ \nonumber
E_{12}^{\ \ \dot{A}\dot{B}} &\equiv& \partial^{(\dot{A}} \big( h^{\dot{B})} - \phi^{2}k_{\dot{S}} \, Q^{\dot{B})\dot{S}} \big) +\phi^{2} \, \frac{\partial k^{(\dot{A}}}{\partial q_{\dot{B})}} + Q^{\dot{A}\dot{B}} \, \partial^{\dot{S}} (\phi^{2}k_{\dot{S}})=0
\\ \nonumber
E_{22}^{\ \ \dot{A}\dot{B}} &\equiv& 2 \, \eth^{(\dot{A}}h^{\dot{B})} + 2\phi^{2} \, \eth_{\dot{S}}Q^{\dot{S}(\dot{A}} k^{\dot{B})} - 2\phi^{2} \, h^{\dot{S}} \partial_{\dot{S}} Q^{\dot{A}\dot{B}}=0
\\ \nonumber
\frac{1}{2} E &\equiv& 4\chi -\eth^{\dot{N}} k_{\dot{N}} + \partial^{\dot{N}} h_{\dot{N}} + 4\phi^{-1} h_{\dot{S}}J^{\dot{S}} - \phi^{4} k_{\dot{S}} \, \partial_{\dot{A}} (\phi^{-2} Q^{\dot{A}\dot{S}})=0
\ \ \ \ \ \ \ \ \ \ \ \ \ \ \ \ \ \ \ \ \ \ \ \ \  \ \ \ \ \ \ 
\end{eqnarray}
then
\begin{eqnarray}
\label{definicje_spinorow_LAB}
l_{11} &=& \partial^{\dot{N}} k_{\dot{N}}
\\ \nonumber
2 \, l_{12} &=& \partial_{\dot{N}} h^{\dot{N}} + \eth_{\dot{N}} k^{\dot{N}} +2\phi^{-1} \, h_{\dot{N}}J^{\dot{N}} +\phi^{2} k_{\dot{N}} \, \partial_{\dot{S}} Q^{\dot{S}\dot{N}}
\\ \nonumber
l_{22} &=& \eth_{\dot{N}} h^{\dot{N}} +\phi^{2} k_{\dot{N}} \, \eth_{\dot{S}} Q^{\dot{S}\dot{N}} - 2\phi\, h_{\dot{N}} J_{\dot{M}} \, Q^{\dot{N}\dot{M}}
\ \ \ \ \ \ \ \ \ \ \ \ \ \ \ \ \ \ \ \ \ \ \ \ \ \ \ \ \ \ \ \ \ \ \ \ \ \ \ \ \ \ \ \ \ \ \ \ \ \ 
\end{eqnarray}
Integrability conditions $L_{RST}^{\ \ \ \ \; \dot{A}}$
\begin{eqnarray}
-\frac{1}{\sqrt{2}} \, L_{111}^{\ \ \ \; \dot{A}} &\equiv&  \phi^{-3} \, \partial^{\dot{A}} \big( \phi^{3} l_{11} \big) = 0
\\ \nonumber
-\frac{1}{\sqrt{2}} \, L_{112}^{\ \ \ \; \dot{A}} &\equiv&  \partial^{\dot{A}} (l_{12}+\chi) + \phi \, l_{11} \, J_{\dot{S}}Q^{\dot{S}\dot{A}} -k^{\dot{A}} \Big( C^{(3)} - \frac{1}{3}\Lambda \Big)  = 0
\\ \nonumber
-\frac{1}{\sqrt{2}} \, L_{122}^{\ \ \ \; \dot{A}} &\equiv&  \partial^{\dot{A}} l_{22} +3\phi^{-1} \, l_{22} J^{\dot{A}} + 2\phi \, l_{12} J_{\dot{S}}Q^{\dot{S}\dot{A}}  + h^{\dot{A}} \Big( C^{(3)} + \frac{2}{3} \Lambda \Big) - C^{(2)} k^{\dot{A}} +2 \, \eth^{\dot{A}} \chi =0
\\ \nonumber
-\frac{1}{\sqrt{2}} \, L_{211}^{\ \ \ \; \dot{A}} &\equiv&  \eth^{\dot{A}} l_{11} + \phi \, l_{11} \, \partial_{\dot{S}}(\phi Q^{\dot{S}\dot{A}} ) - 2\phi^{-1} \, l_{12} \, J^{\dot{A}} - k^{\dot{A}} \Big( C^{(3)} + \frac{2}{3} \Lambda \Big) -2 \, \partial^{\dot{A}} \chi  = 0
\\ \nonumber
-\frac{1}{\sqrt{2}} \, L_{212}^{\ \ \ \; \dot{A}} &\equiv&  \eth^{\dot{A}} (l_{12} - \chi) - \phi^{-1} \, l_{22} \, J^{\dot{A}} +\phi^{2} \, l_{11} \, \eth_{\dot{S}}Q^{\dot{S}\dot{A}} + h^{\dot{A}} \Big( C^{(3)} - \frac{1}{3} \Lambda \Big) - k^{\dot{A}} C^{(2)}=0
\\ \nonumber
-\frac{1}{\sqrt{2}} \, L_{222}^{\ \ \ \; \dot{A}} &\equiv&  \eth^{\dot{A}} l_{22} -\phi \, l_{22} \, \partial_{\dot{S}}(\phi Q^{\dot{S}\dot{A}} ) + 2 \phi^{2} \, l_{12} \, \eth_{\dot{S}} Q^{\dot{S}\dot{A}}
+ C^{(2)} h^{\dot{A}} - C^{(1)} k^{\dot{A}} =0
\end{eqnarray}
Integrability conditions $M_{ABCD}$
\begin{eqnarray}
\label{wwarunki_MABCD}
M_{1111} &\equiv& 0
\\ \nonumber
M_{1112} &\equiv& -\frac{3}{2} \phi^{-2} \, C^{(3)} \, \partial^{\dot{N}} (\phi^{2} k_{\dot{N}}) = 0
\\ \nonumber
M_{1122} &\equiv& h_{\dot{N}} \, \partial^{\dot{N}} C^{(3)} - k_{\dot{N}} \, \eth^{\dot{N}} C^{(3)} - 2\chi \, C^{(3)} - \phi^{-2} \, C^{(2)} \, \partial^{\dot{N}} (\phi^{2} k_{\dot{N}}) = 0
\\ \nonumber
M_{1222} &\equiv& h_{\dot{N}} \, \partial^{\dot{N}} C^{(2)} - k_{\dot{N}} \, \eth^{\dot{N}} C^{(2)} +C^{(2)} \, \big[ -2\chi -l_{12} + \phi \, k_{\dot{N}} \, \partial_{\dot{S}} (\phi Q^{\dot{S}\dot{N}}) + 3\phi^{-1} h_{\dot{N}}J^{\dot{N}} \big]
\\ \nonumber
&&+\frac{3}{2} C^{(3)} \, \big[l_{22} - 2\phi^{2} k_{\dot{N}} \cdot \eth_{\dot{S}}Q^{\dot{S}\dot{N}} + 2 \phi \, h_{\dot{N}} J_{\dot{S}} Q^{\dot{N}\dot{S}} \big] - \frac{1}{2} \phi^{-2} C^{(1)} \, \partial^{\dot{N}} (\phi^{2} k_{\dot{N}}) = 0
\\ \nonumber
M_{2222} &\equiv& h_{\dot{N}} \, \partial^{\dot{N}} C^{(1)} - k_{\dot{N}} \, \eth^{\dot{N}} C^{(1)}
-2C^{(2)} \big[ 2 \phi^{2} \, k_{\dot{N}} \, \eth_{\dot{S}} Q^{\dot{S}\dot{N}} - l_{22} -2\phi \, h_{\dot{N}} J_{\dot{S}} Q^{\dot{N}\dot{S}} \big]
\\ \nonumber
&& +2C^{(1)} \, \big[ \phi k_{\dot{N}} \, \partial_{\dot{S}} (\phi Q^{\dot{N}\dot{S}}) + 3\phi^{-1} J^{\dot{N}}h_{\dot{N}} - l_{12} - \chi \big]=0
\end{eqnarray}
Integrability conditions $N_{AB}^{\ \ \ \dot{A}\dot{B}}$
\begin{eqnarray}
N_{11}^{\ \ \dot{A}\dot{B}} &\equiv& 2 \phi^{-2} \, \partial^{(\dot{A}} \big[\phi^{2} \partial^{\dot{B})} \chi \big] = 0
\\ \nonumber
N_{12}^{\ \ \dot{A}\dot{B}} &\equiv& N_{21}^{\ \ \dot{B}\dot{A}} 
\\ \nonumber
N_{21}^{\ \ \dot{A}\dot{B}} &\equiv& 2\phi^{2} \, \frac{\partial }{\partial q_{\dot{A}}} \partial^{\dot{B}} \chi 
 - 2 \phi \, J^{\dot{A}} \, \frac{\partial \chi}{\partial q_{\dot{B}}} - 2\phi \, J^{\dot{A}} \, \partial_{\dot{S}} \chi \cdot Q^{\dot{B}\dot{S}} + \frac{2}{3} \Lambda \chi  \in^{\dot{A}\dot{B}}
\\ \nonumber
&&+ 2\phi^{2} \, \partial_{\dot{S}} \chi \cdot \partial^{\dot{B}} Q^{\dot{S}\dot{A}}
  + 2\phi \, J_{\dot{S}} \, \partial^{\dot{A}} \chi \cdot Q^{\dot{B}\dot{S}} 
+ 2\phi^{2} \, Q^{\dot{A}\dot{S}} \, \partial_{\dot{S}} \partial^{\dot{B}} \chi =0
\\ \nonumber
N_{22}^{\ \ \dot{A}\dot{B}} &\equiv& 2 \, \eth^{\dot{A}} \eth^{\dot{B}} \chi +2\phi^{2} \, \partial^{\dot{B}} \chi \cdot \eth_{\dot{S}} Q^{\dot{S}\dot{A}} - 2\phi^{2} \, \eth^{\dot{S}} \chi \cdot \partial_{\dot{S}} Q^{\dot{A}\dot{B}} + 2\phi \in^{\dot{A}\dot{B}} J_{\dot{S}} \, \eth_{\dot{C}} \chi \cdot Q^{\dot{C}\dot{S}}=0
\end{eqnarray}
Integrability conditions $R_{ABC}^{\ \ \ \ \ \dot{A}}$
\begin{eqnarray}
R_{111}^{\ \ \ \dot{A}} &\equiv& 0
\\ \nonumber
R_{112}^{\ \ \ \dot{A}} &\equiv& \frac{\sqrt{2}}{2} \, C^{(3)} \, \partial^{\dot{A}} \chi =0
\\ \nonumber
R_{122}^{\ \ \ \dot{A}} &\equiv& \frac{\sqrt{2}}{2} \Big( C^{(2)} \, \partial^{\dot{A}} \chi -C^{(3)} \, \eth^{\dot{A}} \chi \Big) =0
\\ \nonumber
R_{222}^{\ \ \ \dot{A}} &\equiv& \frac{\sqrt{2}}{2} \Big( C^{(1)} \, \partial^{\dot{A}} \chi -C^{(2)}  \, \eth^{\dot{A}} \chi \Big) =0
\ \ \ \ \ \ \ \ \ \ \ \ \ \ \ \ \ \ \ \ \ \ \ \ \ \ \ \ \ \ \ \ \ \ \ \ \ \ \ \ \ \ \ \ \ \ \ \ \ \ \ \
\end{eqnarray}

\subsection{Preparatory analysis.}

Simple analysis of integrability conditions $R_{ABC}^{\ \ \ \ \ \dot{A}}$ and $N_{AB}^{\ \ \ \dot{A}\dot{B}}$ shows that our problem can be divided into two subcases. With $|C^{(3)}|+ |C^{(2)}|+ |C^{(1)}| \ne 0$ assumed, from the conditions $R_{ABC}^{\ \ \ \ \ \dot{A}}$ it follows that $\partial^{\dot{A}} \chi =0$, so conformal factor is a function of $q^{\dot{M}}$ only. However, inserting it to the $N_{21}^{\ \ \dot{A}\dot{B}}$ one obtains
\begin{equation}
- 2 \phi \, J^{\dot{A}} \, \frac{\partial \chi}{\partial q_{\dot{B}}}  + \frac{2}{3} \Lambda \chi  \in^{\dot{A}\dot{B}} = 0
\end{equation}
It is linear polynomial in $\phi$, so $\Lambda \chi =0 $ and $\frac{\partial \chi}{\partial q^{\dot{A}}}=0 \ \ \rightarrow \ \ \chi=\chi_{0} = \textrm{const}$. Thus we arrive at the conclusion that 
\begin{equation}
\label{podstawowy_warunek_calkowalnosci}
\Lambda  \chi_{0} = 0
\end{equation}
Therefore, four-dimensional space of the types $\textrm{[II,D, III or N]}  \otimes \textrm{[any]}$ with nonzero complex expansion does not admit any conformal Killing symmetries with nonconstant $\chi$. This proves that the example from our previous work \cite{biblio_51} describes the less degenerate space which admits conformal Killing vector with $\chi \ne \textrm{const}$ at all. Moreover, from (\ref{podstawowy_warunek_calkowalnosci}) one infers that for any homothetic Killing symmetry ($\chi_{0} \ne 0$) cosmological constant $\Lambda=0$.

On the other hand, when $C_{ABCD}=0$, the integrability conditions $R_{ABC}^{\ \ \ \ \ \dot{A}}$ are identically satisfied and information about conformal factor $\chi$ must be extracted from $N_{AB}^{\ \ \ \dot{A}\dot{B}}$ conditions. The conformal factor must satisfy the condition $N_{11}^{\ \ \dot{A}\dot{B}}$ which solution is very simple
\begin{equation}
\label{conforemna_postac_czynnika_chi}
\chi = \mathfrak{a} \, \frac{\eta}{\phi} + \mathfrak{b} \, \frac{1}{\phi} + \mathfrak{c}
\end{equation}
where $\mathfrak{a}$, $\mathfrak{b}$ and $\mathfrak{c}$ are arbitrary functions of $q^{\dot{M}}$ only. As we mentioned earlier, we do not enter this subcase here and we deal with it in the next work.

\subsection{Final results.}

Let's summarize the most important results. There are 5 arbitrary functions of $q^{\dot{M}}$ only, $\delta^{\dot{B}}$,  $\epsilon$, $\alpha$ and $\beta$. Components of the homothetic Killing vector are
\begin{subequations}
\begin{eqnarray}
k^{\dot{B}} &=& \phi^{-2} \delta^{\dot{B}}
\\ 
\label{postac_h}
h^{\dot{B}} &=& \delta_{\dot{S}} \, Q^{\dot{S}\dot{B}} + \bigg( 2 \chi_{0} + \frac{2}{\tau} K_{\dot{S}} J_{\dot{N}} \frac{\partial \delta^{\dot{S}}}{\partial q_{\dot{N}}} \bigg) p^{\dot{B}} + \frac{\partial \delta_{\dot{S}}}{\partial q_{\dot{B}}} \, p^{\dot{S}} + \epsilon J^{\dot{B}}
\end{eqnarray}
\end{subequations}
In the base $\Big( \frac{\partial}{\partial q^{\dot{A}}}, \frac{\partial}{\partial p^{\dot{B}}} \Big)$ the homothetic Killing vector
\begin{equation}
K= \delta^{\dot{B}} \frac{\partial}{\partial q^{\dot{B}}} - \bigg[  \bigg( 2 \chi_{0} + \frac{2}{\tau} K_{\dot{S}} J_{\dot{N}} \frac{\partial \delta^{\dot{S}}}{\partial q_{\dot{N}}} \bigg) p^{\dot{B}} + \frac{\partial \delta_{\dot{S}}}{\partial q_{\dot{B}}} \, p^{\dot{S}} + \epsilon J^{\dot{B}}  \bigg] \frac{\partial}{\partial p^{\dot{B}}}
\end{equation}
The system of ten homothetic Killing equations can be reduced to one, \textsl{expanding master equation}
\begin{equation}
\label{expanding_master_equation}
\pounds_{K} W = -W \bigg( 4 \chi_{0} + \frac{4}{\tau} K_{(\dot{S}}J_{\dot{N})} \, \frac{\partial \delta^{\dot{S}}}{\partial q_{\dot{N}}} + \frac{1}{\tau} K_{\dot{N}}J_{\dot{S}} \, \frac{\partial \delta^{\dot{S}}}{\partial q_{\dot{N}}} - \frac{\partial \delta^{\dot{N}}}{\partial q^{\dot{N}}} \bigg) + \mathcal{P}
\end{equation}
where $\pounds_{K} W=KW$ is the Lie derivative of the key function and $\mathcal{P}$ is a fourth order polynomial in $p^{\dot{S}}$
\begin{eqnarray}
\mathcal{P} &:=& - \frac{1}{2 \tau^{3}} \, K_{\dot{S}} K_{\dot{N}} \, \frac{\partial \delta^{\dot{S}}}{\partial q_{\dot{N}}} \, \eta \bigg( \mu \phi^{3} - \frac{\Lambda}{3} \bigg) + \frac{1}{2} \frac{\partial^{2} \delta^{\dot{S}}}{\partial q_{\dot{B}} \partial q_{\dot{R}}} \, p_{\dot{R}} p_{\dot{S}} p_{\dot{B}} \phi^{-1} \\ \nonumber
&&+\alpha \, \phi^{3} + \frac{1}{2} \frac{\partial \epsilon}{\partial q^{\dot{A}}} \, p^{\dot{A}} + \beta
\ \ \ \ \ \ \ \ \ \ \ \ \ \ \ \ \ \ \ \ \ \ \ \ 
\end{eqnarray}
The integrability conditions of homothetic Killing equations read
\begin{subequations}
\begin{eqnarray}
\label{integrability_condition_1}
&& \Lambda \chi_{0} = 0 \ \ \ \ \ \ \chi_{0} = \textrm{const}
\\ 
\label{integrability_condition_2}
&& J_{\dot{A}} J_{\dot{B}} \, \frac{\partial \delta^{\dot{A}}}{\partial q_{\dot{B}}} = 0
\\ 
\label{integrability_condition_3}
&& \delta^{\dot{N}} \frac{\partial \mu}{\partial q^{\dot{N}}} - 4 \mu \chi_{0} - \frac{3}{\tau} \mu \, K_{\dot{S}} J_{\dot{N}} \, \frac{\partial \delta^{\dot{S}}}{\partial q_{\dot{N}}} = 0
\\
\label{integrability_condition_4}
&& \tau \delta^{\dot{N}}  \frac{\partial \nu}{\partial q^{\dot{N}}} - 3 \mu \, \frac{\partial \epsilon}{\partial q^{\dot{N}}} J^{\dot{N}} - 2\epsilon \, \frac{\partial \mu}{\partial q^{\dot{N}}} J^{\dot{N}}
+\tau \nu \bigg( \frac{\partial \delta^{\dot{N}}}{\partial q^{\dot{N}}} - 2 \chi_{0} - \frac{1}{\tau} \, J_{\dot{N}} K_{\dot{S}} \, \frac{\partial \delta^{\dot{S}}}{\partial q_{\dot{N}}} \bigg) = 0
\ \ \ \ \ \ \ \ \ \ \ \ \ \ \ 
\\
\label{integrability_condition_5}
&&6 \mu  \beta + \tau \nu \epsilon + 2 \Lambda \alpha + 2  \delta^{\dot{N}} \frac{\partial \gamma}{\partial q^{\dot{N}}} + \frac{4}{\tau}  \gamma \, J_{\dot{N}} K_{\dot{S}} \frac{\partial \delta^{\dot{N}}}{\partial q_{\dot{S}}}  
\\ \nonumber
&& \ \ \ \ \ \ \ \ \ \ \ \ \ \ \ \ \ \ \ \ \ \ \ \ \ \ \ \ \ \ \ \ \ \ \ \
- \frac{1}{\tau^{3}}  K_{\dot{A}}K_{\dot{B}}K_{\dot{C}} J_{\dot{N}} \, \frac{\partial^{3} \delta^{\dot{N}}}{\partial q_{\dot{A}} \partial q_{\dot{B}} \partial q_{\dot{C}}} = 0 
\\
\label{integrability_condition_6}
&&\mu \, \bigg\{ \delta^{\dot{N}} \frac{\partial \varkappa}{\partial q^{\dot{N}}} + 6  J^{\dot{N}} \frac{\partial \alpha}{\partial q^{\dot{N}}} +  \frac{\nu}{\tau} \, K_{\dot{N}}K_{\dot{S}} \frac{\partial \delta^{\dot{S}}}{\partial q_{\dot{N}}} - \frac{\epsilon}{\tau} \, K^{\dot{N}} \frac{\partial \mu}{\partial q^{\dot{N}}}  \ \ \ \ \ \ \ \ \ \ \ \ \ \ \ 
\\ \nonumber
&& \ \ \ \ \ \ \ \ \ \ \ \ \ \ \ \  \ \ \ \ \ \ \ \ \ \ \ \ \ \ \ 
+ \varkappa \bigg(2 \, \frac{\partial \delta^{\dot{N}}}{\partial q^{\dot{N}}} + \frac{1}{\tau} J_{\dot{N}} K_{\dot{S}} \frac{\partial \delta^{\dot{S}}}{\partial q_{\dot{N}}} - 2 \chi_{0} \bigg) \bigg\} =0 \ \ \ \ \ \ \ \ \ \ \ 
\\ 
\label{integrability_condition_7}
&&\Lambda \, \bigg\{ \delta^{\dot{N}} \frac{\partial \varkappa}{\partial q^{\dot{N}}} + 6  J^{\dot{N}} \frac{\partial \alpha}{\partial q^{\dot{N}}} +  \frac{\nu}{\tau} \, K_{\dot{N}}K_{\dot{S}} \frac{\partial \delta^{\dot{S}}}{\partial q_{\dot{N}}} - \frac{\epsilon}{\tau} \, K^{\dot{N}} \frac{\partial \mu}{\partial q^{\dot{N}}}  \ \ \ \ \ \ \ \ \ \ \ \ \ \ \ 
\\ \nonumber
&& \ \ \ \ \ \ \ \ \ \ \ \ \ \ \ \  \ \ \ \ \ \ \ \ \ \ \ \ \ \ \ 
+ \varkappa \bigg(2 \, \frac{\partial \delta^{\dot{N}}}{\partial q^{\dot{N}}} + \frac{1}{\tau} J_{\dot{N}} K_{\dot{S}} \frac{\partial \delta^{\dot{S}}}{\partial q_{\dot{N}}} - 2 \chi_{0} \bigg) \bigg\} =0 \ \ \ \ \ \ \ \ \ \ \ 
\end{eqnarray}
\end{subequations}
For completeness we give the formulas for the spinors $l_{AB}$
\begin{subequations}
\begin{eqnarray}
\label{postac_l11}
l_{11} &=& -2\phi^{-3} \, J_{\dot{N}}\delta^{\dot{N}}
\\
\label{postac_l12}
l_{12} &=& 2\phi^{-1} \, J^{\dot{N}} \delta_{\dot{N}} \, J_{\dot{B}} \partial^{\dot{B}} W + \phi^{-1} \, K^{\dot{N}}\delta_{\dot{N}} \, \frac{1}{\tau} \bigg(\mu \phi^{3} - \frac{\Lambda}{3} \bigg) + \frac{1}{\tau} \, J_{\dot{N}} K_{\dot{A}} \, \frac{\partial \delta^{\dot{N}}}{\partial q_{\dot{A}}}
\\ 
\label{postac_l22}
l_{22} &=& 2\phi^{3} J_{\dot{N}}\delta^{\dot{N}} \bigg( \gamma - \phi^{-2} (J_{\dot{S}}\partial^{\dot{S}}W)^{2} - \mu\phi^{4} \, \partial_{\phi}(\phi^{-3}W) + \frac{\Lambda}{3} \phi^{-2} \, \partial_{\phi}W \bigg)
\\ \nonumber
&&-2\phi^{2} \frac{1}{\tau} J^{\dot{N}}K^{\dot{S}} \, \frac{\partial \delta_{\dot{N}}}{\partial q^{\dot{S}}} \, J_{\dot{B}}\partial^{\dot{B}}W - \phi^{3} \, \delta^{\dot{N}}p_{\dot{N}}\, \frac{\partial \mu}{\partial q^{\dot{C}}}p^{\dot{C}} + 2\phi^{3} \, \delta^{\dot{N}}p_{\dot{N}} \, N_{\dot{S}}J^{\dot{S}}
\\ \nonumber
&& - \frac{\Lambda}{3 \tau^{2}} \phi \frac{\partial \delta^{\dot{S}}}{\partial q_{\dot{N}}} \Big( 2\phi \, K_{\dot{N}}K_{\dot{S}} + \eta \, (2K_{\dot{S}}J_{\dot{N}}+K_{\dot{N}}J_{\dot{S}}) \Big) - \epsilon \phi \bigg( 2 \mu \phi^{3}+ \frac{\Lambda}{3} \bigg)
\\ \nonumber
&&-\mu \phi^{4} \frac{1}{\tau^{2}} \frac{\partial \delta^{\dot{S}}}{\partial q_{\dot{N}}} \Big( \phi \, K_{\dot{N}}K_{\dot{S}} + \eta (2K_{\dot{N}}J_{\dot{S}} + K_{\dot{S}}J_{\dot{N}}) \Big)
-\phi^{3} \frac{1}{\tau^{2}} K_{\dot{S}}K_{\dot{A}}J_{\dot{N}} \frac{\partial^{2} \delta^{\dot{N}}}{\partial q_{\dot{A}} \partial q_{\dot{S}}}
\end{eqnarray}
\end{subequations}

\subsection{Transformation formulas.}

Transformation formulas for the functions $\delta^{\dot{A}}$, $\epsilon$, $\alpha$ and $\beta$ are useful in analysis of classification of the Killing vector. From the transformation rule of any Killing vector we easily find that
\begin{eqnarray}
\delta'^{\dot{A}} &=& \Delta D^{-1 \ \ \dot{A}}_{\ \ \; \dot{B}}  \, \delta^{\dot{B}}
\\ \nonumber
R'^{\dot{A}} &=& \lambda^{-1} D^{-1 \ \ \dot{A}}_{\ \ \; \dot{B}} \, R^{\dot{B}} - \delta^{\dot{B}} \, \frac{\partial p'^{\dot{A}}}{\partial q^{\dot{B}}}
\end{eqnarray}
where $R^{\dot{A}}$ is defined by (\ref{definicja_era}).
\newline
Using the decomposition
\begin{equation}
\tau \delta^{\dot{A}} = a  K^{\dot{A}} + b J^{\dot{A}} \ \ \ \leftrightarrow \ \ \ a:= J_{\dot{A}} \delta^{\dot{A}} \ \ \ , \ \ \ b:= -K_{\dot{A}} \delta^{\dot{A}}
\end{equation}
where $b=b(w,t)$ and by (\ref{integrability_condition_2}), $a=a(w)$ one gets
\begin{eqnarray}
\label{transformacja_a}
a' &=& w'_{w} a
\\
\label{transformacja_b}
b' &=& \lambda^{-\frac{1}{2}} b + t'_{w} a
\\
\label{transformacja_epsilon}
\epsilon' &=& (\lambda w'_{w})^{-1} \epsilon - \sigma \Big[ 2 \chi_{0} +a_{w} -2b_{t} + a \, ( \ln \sigma \lambda w'_{w})_{w} + b \, (\ln \sigma \lambda)_{t} \Big]
\end{eqnarray}
Much more complicated transformation rules for $\alpha$ and $\beta$ follow from invariancy of the master equation and they read
\begin{eqnarray}
\label{transformacja_alpha}
(w'_{w})^{2} \alpha' &=& \alpha - \frac{1}{2 \tau} \mu \sigma \lambda^{\frac{3}{2}} w'_{w} \Big( \partial_{w} - \frac{t'_{w}}{\lambda^{-\frac{1}{2}}} \, \partial_{t} \Big) \big( \lambda^{-\frac{1}{2}} b + t'_{w} a \big) 
\\ \nonumber
&& - \frac{1}{2 \tau} \epsilon \mu \lambda^{\frac{1}{2}} t'_{w}  - \frac{1}{3} a L_{w} - \frac{1}{3} b L_{t} + \frac{2}{3} L (\chi_{0} - a_{w})
\\
\label{transformacja_beta}
(w'_{w})^{2} \lambda^{\frac{3}{2}} \beta' &=& \beta -\frac{1}{2} \tau \sigma \lambda^{2} (w'_{w})^{2} \frac{\partial \epsilon'}{\partial t} + \frac{1}{2} \tau (\sigma \lambda w'_{w})^{2} \partial_{t} \Big( \lambda^{\frac{1}{2}} \, \partial_{t} \big(  \lambda^{-\frac{1}{2}} b + t'_{w} a \big) \Big)
\\ \nonumber
&& +\frac{1}{2}\tau \lambda \epsilon \sigma_{t} \, w'_{w} -M (4\chi_{0} -3 b_{t} +2 a_{w} ) - a M_{w} - b M_{t} 
\\ \nonumber
&& +\frac{\Lambda}{6 \tau} \epsilon \lambda^{\frac{1}{2}} t'_{w} + \frac{\Lambda}{6 \tau}\sigma w'_{w} \lambda^{\frac{3}{2}} \Big(\partial_{w} - \frac{t'_{w}}{\lambda^{-\frac{1}{2}}} \, \partial_{t} \Big) \big(  \lambda^{-\frac{1}{2}} b + t'_{w} a \big)
\end{eqnarray}
The formulae from (\ref{transformacja_a}) to (\ref{transformacja_beta}) are presented in the form very similar to that from Ref. \cite{biblio_52}. It seems, that there are several small misprints in transformation formulas in \cite{biblio_52}.

It is useful to present integrability conditions (\ref{integrability_condition_2}) - (\ref{integrability_condition_7}), the master equation and the form of the homothetic Killing vector, by using $a$, $b$ and coordinates $(\phi, \eta, w, t)$ instead of $\delta^{\dot{A}}$ and $(p^{\dot{A}}, q^{\dot{B}})$. Thus we have
\begin{subequations}
\begin{eqnarray}
&& a\mu_{w}+b \mu_{t} - 4  \mu \chi_{0}+ 3\mu b_{t} = 0
\\
&& a\nu_{w}+b \nu_{t} - 3 \mu \epsilon_{t} - 2 \epsilon \mu_{t} + \nu  \big( a_{w}+2b_{t}-2\chi_{0} \big)=0
\\
&&6\mu \beta +\tau \nu \epsilon + 2 \Lambda \alpha + 2 a \gamma_{w} +2 b \gamma_{t} + 4 \gamma a_{w} - a_{www}=0
\\
&&\mu \Big( a \varkappa_{w} + b \varkappa_{t} + 6 \tau \alpha_{t} - \nu b_{w} - \epsilon \mu_{w} + \varkappa \big(2 a_{w} + b_{t} -2\chi_{0} \big) \Big)=0
\\
&&\Lambda \Big( a \varkappa_{w} + b \varkappa_{t} + 6 \tau \alpha_{t} - \nu b_{w} - \epsilon \mu_{w} + \varkappa \big(2 a_{w} + b_{t} -2\chi_{0} \big) \Big)=0
\end{eqnarray}
\end{subequations}
and the form of the homothetic Killing vector
\begin{eqnarray}
K&=&a \, \frac{\partial}{\partial w} + b \, \frac{\partial}{\partial t} +  (b_{t}-2\chi_{0}) \phi   \, \frac{\partial}{\partial \phi} 
\\ \nonumber
&&+ \Big( (2b_{t}-a_{w} -2 \chi_{0} ) \eta + b_{w} \phi - \tau \epsilon \Big) \frac{\partial}{\partial \eta}
\end{eqnarray}
Finally, the master equation reads
\begin{eqnarray}
\pounds_{K} W &=& - (4 \chi_{0} +2a_{w} - 3b_{t} ) W + \frac{b_{w}}{2 \tau^{2}} \eta \Big( \mu \phi^{3} - \frac{\Lambda}{3} \Big) + \alpha \phi^{3}
\\ \nonumber
&& +\frac{1}{2 \tau} \Big( -b_{ww} \phi^{2} - b_{tt} \eta^{2} + (a_{ww} - 2 b_{tw}) \eta \phi \Big) 
+ \frac{1}{2} ( \epsilon_{w} \phi + \epsilon_{t} \eta) + \beta
\end{eqnarray}

\renewcommand{\arraystretch}{1.2}
\setlength\arraycolsep{2pt}
\setcounter{equation}{0}

\section{Classification of Killing vectors.}

Classification of Killing vectors can be done in many different ways. Algebraic type of the self-dual conformal curvature can be treated as a main criterion and then the form of the covariant derivative of the Killing vector described by the spinor $l_{AB}$ is the subcriterion. That way was chosen in \cite{biblio_52}. We propose the different way, presented in the table:
\newline
\newline
\begin{tabular}{|c|c|c|c|}   \hline
Isometric symmetries & Functions   &  Homothetic symmetries  & Functions \\  \hline
Type IK1 &   $a \ne0$                   &  Type HK1   &   $a \ne0$   \\ \hline
Type IK2 &   $a=0$, $b\ne0$             &  Type HK2   &   $a=0$, $b\ne0$  \\ \hline
Type IK3 &   $a=b=0$, $\epsilon \ne0$   &  Type HK3   &   $a=b=\epsilon=0$ \\ \hline
\end{tabular}
\newline
\newline
Using the transformation formulas (\ref{transformacja_a})-(\ref{transformacja_beta}) we can reduce the form of the Killing vector to one of three canonical forms, namely $K=\partial_{w}$, $K=\partial_{t}$ or $K=\partial_{\eta}$ (note the difference with \cite{biblio_52}). The second step is to use the transformation formulas (\ref{trans_mu}) - (\ref{trans_gamma}) and the remaining gauge freedom in order to simplify the hyperheavenly equation. 

Generally the results in the tables below follow from the analysis of the master equation and its integrability conditions. In some cases important constraints on structural functions are given by analysis of the hyperheavenly equation. In order to distinguish those cases, we use the symbol $\stackrel{\mathcal{HH}}{=}$. [For example, in type IK1, algebraic types $[\textrm{III,N}] \otimes [\textrm{any}]$ with $\Lambda=0$ has $\varkappa=\varkappa(w,t)$ and remainig gauge freedom seems to be useless. However, simple analysis of hyperheavenly equation proves, that $\varkappa_{w}=0$, so $\varkappa=\varkappa(t)$ and it can be gauged away. In that case we marked it as $\varkappa \stackrel{\mathcal{HH}}{=} 0$].

\subsection{Isometric Killing symmetries.}

Here we give a classification of isometric Killing vectors (IK). For all subcases $\chi_{0}=0$. When some constant or function cannot be identically equal to $0$, it is clearly marked as $\ne0$. All nonzero constants except cosmological constant, can be re-scaled to $1$.

\subsubsection{Type IK1 ($K=\partial_{w}$)}

The form of Killing vector and the key function:
\newline
\newline
\begin{tabular}{|c|c|c|c|}   \hline
Functions & Killing vector  & Master equation & The key function  \\  \hline
$a=1$,  $b=\epsilon=\alpha=\beta=0$ & $K=\partial_{w}$ & $\pounds_{K} W=0$   & $W=W(\eta, \phi, t)$ 
\\ \hline
\end{tabular}
\newline
\newline
Concrete algebraic types are characterized by:
\newline
\newline
\begin{tabular}{|c|c|c|c|c|}   \hline
$[\textrm{II, D}] \otimes [\textrm{any}]$ & \multicolumn{2}{|c|}{$[\textrm{III}] \otimes [\textrm{any}]$}
& \multicolumn{2}{|c|}{$[\textrm{N}] \otimes [\textrm{any}]$} \\ \hline
$\mu=\mu_{0}\ne0$, $\nu=0$ & \multicolumn{2}{|c|}{$\mu=0$, $\nu=\nu_{0} \ne 0$ } & \multicolumn{2}{|c|}{$\mu=\nu=0$} \\ \cline{2-5}
$\Lambda$ arbitrary & $\Lambda=0$ & $\Lambda \ne0$ & $\Lambda=0$ & $\Lambda \ne0$ \\ 
$\varkappa=0$  & $\varkappa \stackrel{\mathcal{HH}}{=}0$ & $\varkappa=0$ & $\varkappa \stackrel{\mathcal{HH}}{=}0$ & [$\varkappa=\varkappa_{0} \ne 0$, $\gamma=0$] \\ 
$\gamma=0$& $\gamma=0$ & $\gamma=0$ & $\gamma=\gamma(t)$, $\gamma_{t} \ne0$  & or \\ 
 &  &  &   &  [ $\varkappa=0$, $\gamma=\gamma(t)$, $\gamma_{t} \ne 0$]  \\ \hline
\end{tabular}

\subsubsection{Type IK2 ($K=\partial_{t}$)}

The form of Killing vector and the key function:
\newline
\newline
\begin{tabular}{|c|c|c|c|}   \hline
Functions & Killing vector  & Master equation & The key function  \\  \hline
$b=1$,  $a=\epsilon=\alpha=\beta=0$ & $K=\partial_{t}$ & $\pounds_{K} W=0$   & $W=W(\eta, \phi, w)$ 
\\ \hline
\end{tabular}
\newline
\newline
Concrete algebraic types are characterized by:
\newline
\newline
\begin{tabular}{|c|c|c|c|c|c|}   \hline
\multicolumn{3}{|c|}{$[\textrm{II, D}] \otimes [\textrm{any}]$} & \multicolumn{2}{|c|}{$[\textrm{III}] \otimes [\textrm{any}]$}  &  $[\textrm{N}] \otimes [\textrm{any}]$  \\ \hline
$\mu=\mu(w)$, $\mu_{w} \ne 0$  &  \multicolumn{2}{|c|}{$\mu=\mu_{0}\ne0$}  & \multicolumn{2}{|c|}{$\mu=0$ }  &  $\mu=0$  \\ \cline{2-3}
$\nu=\nu_{0}$ & $\nu=\nu_{0} \ne0$  &  $\nu=0$  &  \multicolumn{2}{|c|}{$\nu=\nu_{0}\ne0$}   &  $\nu=0$  \\ \cline{4-5}
$\Lambda$ arbitrary & $\Lambda$ arbitrary  &  $\Lambda$ arbitrary  &  $\Lambda=0$ & $\Lambda\ne 0$   &  $\Lambda\ne 0$  \\
 $\varkappa=0$ & $\varkappa=0$  &  $\varkappa=\varkappa_{0}$  &  $\varkappa \stackrel{\mathcal{HH}}{=} 0$  &  $\varkappa=0$  &  $\varkappa=\varkappa_{0} \ne 0$ \\
$\gamma=0$ & $\gamma=0$  &  $\gamma=0$  &  $\gamma=0$  &  $\gamma=0$  &  $\gamma=0$ \\ \hline
\end{tabular}

\subsubsection{Type IK3 ($K=\partial_{\eta}$)}
\label{subsubsekcja_typ_IK3}

The form of Killing vector and the key function:
\newline
\newline
\begin{tabular}{|c|c|c|c|}   \hline
Functions & Killing vector & Master equation & The key function  \\  \hline
$\epsilon=-\frac{1}{\tau}$,  $a=b=\beta=0$ & $K=\partial_{\eta}$   & $\pounds_{K} W=\alpha \phi^{3}$   & $W=\alpha \phi^{3} \eta + F(\phi, w, t)$ \\ \hline
\end{tabular}
\newline
\newline
Concrete algebraic types are characterized by:
\newline
\newline
\begin{tabular}{|c|c|c|c|}   \hline
$[\textrm{II, D}] \otimes [\textrm{any}]$ & $[\textrm{III}] \otimes [\textrm{any}]$  &  \multicolumn{2}{|c|}{$[\textrm{N}] \otimes [\textrm{any}]$}  \\ \hline
$\mu=\mu_{0} \ne 0$, $\nu=0$  &  $\mu=0$, $\nu=2\alpha_{0}\Lambda\ne 0$  &  \multicolumn{2}{|c|}{$\mu=\nu=0$}  \\ \cline{3-4}
$\Lambda$ arbitrary  &  $\Lambda \ne 0$  &  $\Lambda =0$  &  $\Lambda \ne 0$  \\
$\alpha=0$ &  $\alpha=\alpha_{0}\ne0$  &  $\alpha \stackrel{\mathcal{HH}}{=} \alpha_{0}$  &  $\alpha=0$  \\
$\varkappa=\gamma=0$  &  [$\varkappa=0$, $\gamma=\gamma(w,t)$]  &  $\varkappa=0$  &  [$\varkappa=0$, $\gamma=\gamma(w,t)$, $\gamma_{t} \ne 0$]  \\
  &  or  &  $\gamma=\gamma(w,t)$, $\gamma_{t} \ne 0$  &  or  \\
 &  [$\varkappa=\varkappa(w,t)$, $\gamma=0$]  &  & [$\varkappa=\varkappa(w,t) \ne 0$, $\gamma=0$] \\ \hline 
\end{tabular}

\subsection{Homothetic Killing symmetries with $\chi_{0} \ne 0$.}

In this subsection complete classification of homothetic Killing vectors (HK) with $\chi_{0} \ne 0$ is done. 
In all subcases $\chi_{0} \ne 0$, but cosmological constant $\Lambda=0$.

\subsubsection{Type HK1 ($K=\partial_{w} - 2\chi_{0} (\phi \partial_{\phi} + \eta \partial_{\eta})$)}

The form of Killing vector and the key function:
\newline
\newline
\begin{tabular}{|c|c|c|c|}   \hline
Functions & Killing vector  & Master equation & The key function  \\  \hline
$a=1$,  $b=\epsilon=\alpha=\beta=0$ & $K=\partial_{w} - 2\chi_{0} (\phi \partial_{\phi} + \eta \partial_{\eta})$ & $\pounds_{K} W=-4\chi_{0} W$   & $W=e^{-4 \chi_{0} w }F(x,y,t)$ 
\\ \hline
\end{tabular}
\newline
\newline
where $x:=\phi \, e^{2 \chi_{0} w}$ and $y:=\eta \, e^{2 \chi_{0} w}$
\newline
\newline
Concrete algebraic types are characterized by:
\newline
\newline
\begin{tabular}{|c|c|c|}   \hline
$[\textrm{II, D}] \otimes [\textrm{any}]$ & $[\textrm{III}] \otimes [\textrm{any}]$  &  $[\textrm{N}] \otimes [\textrm{any}]$  \\ \hline
$\mu=e^{4\chi_{0} w}$  &  $\mu=0$ & $\mu=0$  \\ 
$\nu=0$  &  $\nu=e^{2\chi_{0}w}$ & $\nu=0$  \\ 
$\varkappa=0$  &  $\varkappa \stackrel{\mathcal{HH}}{=}  0$ & $\varkappa \stackrel{\mathcal{HH}}{=}  0$  \\ 
$\gamma=0$  &  $\gamma=0$ & $\gamma=\gamma(t)$ , $\gamma_{t} \ne 0$ \\ \hline
\end{tabular}

\subsubsection{Type HK2 ($K=\partial_{t} - 2\chi_{0} (\phi \partial_{\phi} + \eta \partial_{\eta})$)}

The form of Killing vector and the key function:
\newline
\newline
\begin{tabular}{|c|c|c|c|}   \hline
Functions & Killing vector  & Master equation & The key function  \\  \hline
$b=1$,  $a=\epsilon=\alpha=\beta=0$ & $K=\partial_{t} - 2\chi_{0} (\phi \partial_{\phi} + \eta \partial_{\eta})$ & $\pounds_{K} W=-4\chi_{0} W$   & $W=e^{-4 \chi_{0} t }F(x,y,w)$ 
\\ \hline
\end{tabular}
\newline
\newline
where $x:=\phi \, e^{2 \chi_{0} t}$ and $y:=\eta \, e^{2 \chi_{0} t}$
\newline
\newline
Concrete algebraic types are characterized by:
\newline
\newline
\begin{tabular}{|c|c|c|}   \hline
$[\textrm{II, D}] \otimes [\textrm{any}]$ & $[\textrm{III}] \otimes [\textrm{any}]$  &  $[\textrm{N}] \otimes [\textrm{any}]$  \\ \hline
$\mu=\kappa(w)e^{4\chi_{0} t}$, $\kappa \ne 0$  &  $\mu=0$ & does not  \\ 
$\nu=0$  &  $\nu=e^{2\chi_{0}t}$ &  admit any \\ 
$\varkappa=0$  &  $\varkappa \stackrel{\mathcal{HH}}{=}  0$ & homothetic vector  \\ 
$\gamma=0$  &  $\gamma=0$ & of the Type HK2 \\ \hline
\end{tabular}

\subsubsection{Type HK3 ($K= - 2\chi_{0} (\phi \partial_{\phi} + \eta \partial_{\eta})$)}

The form of Killing vector and the key function:
\newline
\newline
\begin{tabular}{|c|c|c|c|}   \hline
Functions & Killing vector  & Master equation & The key function  \\  \hline
  $a=b=\epsilon=\alpha=\beta=0$ & $K=- 2\chi_{0} (\phi \partial_{\phi} + \eta \partial_{\eta})$ & $\pounds_{K} W=-4\chi_{0} W$   & $W=W(p^{\dot{A}}, q^{\dot{B}})$ 
\\ \hline
\end{tabular}
\newline
\newline
The master equation gives 
\begin{equation}
 \phi \, \frac{\partial W}{\partial \phi} + \eta \, \frac{\partial W}{\partial \eta}  = 2 W \ \ \ \rightarrow \ \ \ 
p^{\dot{A}} \, \frac{\partial W}{\partial p^{\dot{A}}} = 2W
\end{equation}
so $W$ is an arbitrary homogenous function of the degree 2 in the variables $p^{\dot{A}}$.
\newline
\newline
Concrete algebraic types are characterized by:
\newline
\newline
\begin{tabular}{|c|c|c|}   \hline
$[\textrm{II, D}] \otimes [\textrm{any}]$ & $[\textrm{III}] \otimes [\textrm{any}]$  &  $[\textrm{N}] \otimes [\textrm{any}]$  \\ \hline
does not  &  does not & $\mu=0$  \\ 
admit any &  admit any & $\nu=0$  \\ 
homothetic vector  &  homothetic vector & $\gamma=\gamma(w,t)$, $\gamma_{t} \ne 0$  \\ 
of the Type HK3  &  of the Type HK3 & $\varkappa=\varkappa_{0}$ \\ \hline
\end{tabular}

\setlength\arraycolsep{2pt}
\setcounter{equation}{0}

\section{Examples of expanding $\mathcal{HH}$-spaces with symmetry.}
\label{sekcja_klasyfikacja}

\subsection{Type $[\textrm{D}] \otimes [\textrm{any}]$}

As a first example we consider the algebraic type $[\textrm{D}] \otimes [\textrm{any}]$. We show the way to reduce the hyperheavenly equation as much as possible. Type $[\textrm{D}] \otimes [\textrm{any}]$ appears when 
\begin{equation}
\label{wwarunek_na_typ_D}
2C^{(2)}C^{(2)}-3C^{(1)}C^{(3)}=0
\end{equation}
Equation (\ref{wwarunek_na_typ_D}) is a differential equation for the key function $W$ and this equation can be easily solved in all subcases.
\newline
\textbf{Type IK1}
\newline
Equation (\ref{wwarunek_na_typ_D}) yields $W_{t}=0 \ \ \rightarrow \ \ W=W(\phi, \eta)$ and the hyperheavenly equation reads
\begin{eqnarray}
\label{zredukowane_rroowwnanie_na_typ_D_1}
&&  \tau^{2} \big( W_{\eta \eta}W_{\phi \phi} - W_{\eta \phi}W_{\eta \phi} + 2 \phi^{-1} W_{\eta} W_{\eta \phi} - 2 \phi^{-1} W_{\phi}W_{\eta\eta} \big)
\\ \nonumber
&&- \mu_{0} \big( \phi^{2} W_{\phi\phi} - 3 \phi W_{\phi} + 3 W \big)  - \frac{\Lambda}{6} \phi^{-1} W_{\phi \phi} = 0
\end{eqnarray}
(Note that $\mu_{0}$ can be gauged to $1$).
\newline
\textbf{Type IK2}
\newline
For all three subcases pointed in the respective table we find that the type $[\textrm{D}] \otimes [\textrm{any}]$ condition (\ref{wwarunek_na_typ_D}) is equivalent to the equation
\begin{equation}
\label{warunek_na_typ_D_dla_IK2}
(\tau^{2} \nu_{0} + \tau \mu_{w} \phi) \, \frac{\partial W}{\partial \eta} = \frac{(\tau \nu_{0} + 2 \mu_{w} \phi)^{2}}{6\mu} - \frac{1}{2} \mu_{ww} \phi^{2} 
+  \Big(\mu\phi^{3}- \frac{\Lambda}{3} \Big) \Big( \frac{\mu_{w}\eta}{2\tau} + \frac{\varkappa_{0}}{2} \Big)
\end{equation}
When $\mu_{w}=0$ and $\nu_{0}=0$ the equation (\ref{warunek_na_typ_D_dla_IK2}) gives no constraints on the key function, but $\varkappa_{0}$ must vanish. Then the hyperheavenly equation takes the form
\begin{eqnarray}
&& \tau^{2} \big( W_{\eta \eta}W_{\phi \phi} - W_{\eta \phi}W_{\eta \phi} + 2 \phi^{-1} W_{\eta} W_{\eta \phi} - 2 \phi^{-1} W_{\phi}W_{\eta\eta} \big) +  \tau \phi^{-1}  W_{w\eta}
\\ \nonumber
&&-  \mu_{0} \big( \phi^{2} W_{\phi\phi} - 3 \phi W_{\phi} + 3 W \big)  - \frac{\Lambda}{6} \phi^{-1} W_{\phi \phi} = 0
\end{eqnarray}
with $W=W(\eta, \phi, w)$. Note that $\mu_{0}$ can be gauged to 1.
\newline
When at least one of $\mu_{w}$ or $\nu_{0}$ is $\ne0$ we find the general solution of 
(\ref{warunek_na_typ_D_dla_IK2}) to be (remember that $\varkappa_{0}=0$)
\begin{equation}
\label{funkcja_kluczowa_dla_typu_D_ww_typie_IK2}
W = \frac{(\nu_{0}\tau + 2 \mu_{w} \phi)^{2} - 3 \mu \mu_{ww} \phi^{2}}{6\mu (\tau^{2} \nu_{0} + \tau \mu_{w} \phi)} \eta 
+ \frac{\mu_{w}(\mu \phi^{3} - \frac{\Lambda}{3})}{4 \tau(\tau^{2} \nu_{0} + \tau \mu_{w} \phi)} \eta^{2} + F(\phi, w)
\end{equation}
with $F(\phi, w)$ being an arbitrary function. Inserting this into hyperheavenly equation together with $\varkappa=\gamma=0$, one gets the second-order polynomial in $\eta$. The conditions following from quadratic and linear terms in $\eta$ are identically satisfied. Finally, one arrives at the equation for $F$
\begin{equation}
\label{zredukowane_rroowwnanie_na_typ_D_2}
\Big(2\tau^{2} m \phi - \mu \phi^{3} - \frac{\Lambda}{6} \Big) F_{\phi \phi} + (3\mu \phi^{2} - 4 \tau^{2} m) F_{\phi} - 3\mu \phi F + \tau n_{w} +2\tau^{2} nn_{\phi} - \tau^{2} \phi n_{\phi}^{2} = 0
\end{equation}
where $n$ and $m$ are given by
\begin{equation}
m=\frac{\mu_{w}(\mu \phi^{3} - \frac{\Lambda}{3})}{4 \tau(\tau^{2} \nu_{0} + \tau \mu_{w} \phi)}
\ \ \ \ , \ \ \ \ 
n=\frac{(\nu_{0}\tau + 2 \mu_{w} \phi)^{2} - 3 \mu \mu_{ww} \phi^{2}}{6\mu (\tau^{2} \nu_{0} + \tau \mu_{w} \phi)}
\end{equation}
We are going to study Eqs. (\ref{zredukowane_rroowwnanie_na_typ_D_1}) and (\ref{zredukowane_rroowwnanie_na_typ_D_2}) elsewhere.
\newline
[It is easy to check (by substituting (\ref{funkcja_kluczowa_dla_typu_D_ww_typie_IK2}) into the master equation), that in general the second Killing vector does not exist].
\newline
\textbf{Type IK3}
\newline
Equation (\ref{wwarunek_na_typ_D}) gives $F_{t}=0 \ \ \rightarrow \ \ F=F(\phi, w)$ and consequently the hyperheavenly equation reduces to
\begin{equation}
 \mu_{0} \big( \phi^{2} F_{\phi\phi} - 3 \phi F_{\phi} + 3 F \big)  + \frac{\Lambda}{6} \phi^{-1} F_{\phi \phi} = 0
\end{equation}
($\mu_{0}$ can be gauged to 1).
\newline
This equation can be easily solved (see subsection \ref{subsekcja_na_typy}).
\newline
\textbf{Type HK1}
\newline
The hyperheavenly equation for the types $[\textrm{II, D}] \otimes [\textrm{any}]$ in $(x,y,w,t)$ coordinates reads
\begin{eqnarray}
\label{rownanie hiperniebianskie_na_typ_D2_homotetyczny1}
&&\tau^{2} \Big( F_{yy}F_{xx}-F_{xy}^{2} + 2 x^{-1} F_{y}F_{xy} - 2 x^{-1} F_{x} F_{yy} \Big) 
\\ \nonumber
&&+ \tau x^{-1} \Big( 2 \chi_{0} (yF_{yy}+xF_{xy} -F_{y}) -F_{xt} \Big) - \Big( x^{2} F_{xx} - 3x F_{x} +3F \Big) - \frac{2 \chi_{0}}{\tau} xy = 0
\end{eqnarray}
where $F=F(x,y,t)$ is an arbitrary function of its variables, and $x:=\phi \, e^{2 \chi_{0} w}$, $y:=\eta \, e^{2 \chi_{0} w}$. 
\newline
Condition (\ref{wwarunek_na_typ_D}) for the type $[\textrm{D}] \otimes [\textrm{any}]$ implies the differential equation of the general solution 
\begin{equation}
\label{funkcja_kluczowa_na_typ_D_homotetyczny1}
F(x,y,t) = \frac{1  }{4 \tau^{2} } \, x^{2}y^{2} + \frac{2 \chi_{0} }{3 \tau } \, yx  + f \Big(x, y+ \frac{4}{3}\chi_{0} xt \Big)
\end{equation}
with $f$ being an arbitrary function of its variables modulo the hyperheavenly equation. Inserting (\ref{funkcja_kluczowa_na_typ_D_homotetyczny1}) into (\ref{rownanie hiperniebianskie_na_typ_D2_homotetyczny1}) we obtain
\begin{eqnarray}
&&\tau^{2} \big(f_{xx}f_{zz}-f_{xz}^{2} + 2 x^{-1} f_{z}f_{xz} - 2 x^{-1} f_{x}f_{zz} \big) -\frac{1}{2} \big( x^{2} f_{xx} + 2xz f_{xz} +z^{2} f_{zz} \big)
\\ \nonumber
&&+2 xf_{x} + 2zf_{z} + \frac{2}{3} \tau \chi_{0} x^{-1} (z f_{zz} + x f_{xz} ) - 2 \tau \chi_{0} x^{-1} f_{z} -3f  + \frac{4}{9} \chi_{0}^{2} =0 
\end{eqnarray}
where $f=f(x,z)$ and $z$ stands for $z=y + \frac{4}{3} \chi_{0} xt$.
\newline
\textbf{Type HK2}
\newline
Now, the hyperheavenly equation for the types $[\textrm{II, D}] \otimes [\textrm{any}]$ reads
\begin{eqnarray}
\label{rownanie hiperniebianskie_na_typ_D2_homotetyczny2}
&&\tau^{2} \Big( F_{yy}F_{xx}-F_{xy}^{2} + 2 x^{-1} F_{y}F_{xy} - 2 x^{-1} F_{x} F_{yy} \Big) 
\\ \nonumber
&&+ \tau x^{-1} \Big( F_{yw} - 2\chi_{0} (xF_{xx} + yF_{xy} -F_{x}) \Big)
- \kappa \Big( x^{2} F_{xx} - 3x F_{x} +3F \Big) 
\\ \nonumber
&&+\frac{1}{2 \tau} \Big( 4 \chi_{0} \kappa y^{2} - \kappa_{w} xy \Big) = 0
\end{eqnarray}
where $F=F(x,y,w)$ is an arbitrary function of its variables and $x:=\phi \, e^{2 \chi_{0} t}$, $y:=\eta \, e^{2 \chi_{0} t}$. Then the condition (\ref{wwarunek_na_typ_D}) determining the type $[\textrm{D}] \otimes [\textrm{any}]$ leads to
\begin{eqnarray}
\label{funkcja_kluczowa_na_typ_D_homotetyczny2}
F(x,y,w) &=& - \frac{\kappa_{w}^{2}}{64 \tau^{2} \chi_{0}^{2} \kappa} \, x^{4} - \frac{\kappa_{w}}{8 \tau^{2} \chi_{0}} \, yx^{3} + \frac{3 \kappa \kappa_{ww}-4 \kappa_{w}^{2}}{24 \tau \kappa^{2} \chi_{0}} \, x^{2} + \frac{2\chi_{0}}{3 \tau} \, y^{2} 
\\ \nonumber
&&+ f \Big(xy+\frac{\kappa_{w}}{4\kappa\chi_{0}}x^{2},w \Big)
\end{eqnarray}
with $f$ beeing the arbitrary function of its variables. Inserting (\ref{funkcja_kluczowa_na_typ_D_homotetyczny2}) into (\ref{rownanie hiperniebianskie_na_typ_D2_homotetyczny2}) we find the form of the reduced hyperheavenly equation
\begin{eqnarray}
&& \tau^{2} \big( f_{z}^{2} -  2 z f_{z}f_{zz} \big) - \Big( \kappa z^{2} + \frac{\tau \kappa_{w}}{3 \kappa} z \Big) f_{zz} + \Big( 3 \kappa z - \frac{2 \tau \kappa_{w}}{3 \kappa} \Big) f_{z}
\\ \nonumber
&& +\tau f_{zw} - 3 \kappa f -\frac{3 \kappa \kappa_{ww}- 4 \kappa_{w}^{2}}{9 \kappa^{2}}=0
\end{eqnarray}
where $z=xy+\frac{\kappa_{w}}{4\kappa\chi_{0}}x^{2}$ and $f=f(z,w)$.
\newline
\textbf{Type HK3}
\newline
This type does not admit any algebraic type $[\textrm{D}] \otimes [\textrm{any}]$.

\subsection{Type IK3}
\label{subsekcja_na_typy}

As a second example we present some explicit solutions for the type IK3 for different algebraic types. All possible simplifications are gathered in subsection (\ref{subsubsekcja_typ_IK3}).
\newline
\textbf{Type $[\textrm{II,D}] \otimes [\textrm{any}]$}
\newline
We have here $\mu_{0} \ne 0$, $\nu=\varkappa=\gamma=\alpha=0$, $\Lambda$ is arbitrary. The hyperheavenly equation gives
\begin{equation}
\label{zredukowanerownanienatypIId}
\Big(\mu_{0} \phi^{3} + \frac{\Lambda}{6} \Big) F_{\phi \phi} - 3 \mu_{0} \phi^{2} F_{\phi} + 3 \mu_{0} \phi F + \tau F_{\phi t} = 0
\end{equation}
In Eqs. (\ref{zredukowanerownanienatypIId}) the variables can be separated. This example was presented in \cite{biblio_52}.
\newline
\textbf{Type $[\textrm{D}] \otimes [\textrm{any}]$}
\newline
Condition (\ref{wwarunek_na_typ_D}) gives $F_{t}=0$ and this yields
\begin{equation}
\label{zredukowanerownanienatypd}
\Big(\mu_{0} \phi^{3} + \frac{\Lambda}{6} \Big) F_{\phi \phi} - 3 \mu_{0} \phi^{2} F_{\phi} + 3 \mu_{0} \phi F  = 0
\end{equation}
The general solution of (\ref{zredukowanerownanienatypd}) is
\begin{equation}
F(\phi, w) = \mu_{0} \, \mathfrak{f} (w) \, \phi^{3} + \mathfrak{g} (w) \, \phi - \frac{\Lambda}{3 } \, \mathfrak{f} (w)
\end{equation}
where $\mathfrak{f}$ and $\mathfrak{g}$ are arbitrary functions of their variable. The metric reads
\begin{equation}
ds^{2} = (\phi \tau)^{-2} \bigg\{ 2\tau (d \eta \underset{s}{\otimes} d w - d \phi \underset{s}{\otimes} dt) + 2 \, \Big(  \mu_{0} \phi^{3}+\frac{\Lambda}{6}  \Big) 
dt \underset{s}{\otimes}dt + 2 \tau^{2} \mathfrak{g} \, dw \underset{s}{\otimes}dw \bigg\} 
\end{equation}
Curvature
\begin{equation}
C^{(3)}=- 2 \mu_{0} \phi^{3} \ \ \ , \ \ \ C^{(2)}=C^{(1)}=0 \ \ \ , \ \ \ C_{\dot{A}\dot{B}\dot{C}\dot{D}} = - \frac{6 \mu_{0}}{\tau^{2}} \phi^{3} \, J_{(\dot{A}}J_{\dot{B}}K_{\dot{C}}K_{\dot{D})}
\end{equation}
The form of $C_{\dot{A}\dot{B}\dot{C}\dot{D}}$ proves that now only the type $[\textrm{D}] \otimes [\textrm{D}]$ is allowed. 
\newline
(In all formulas $\mu_{0}$ can be gauged to $1$, if desired).
\newline
\textbf{Type $[\textrm{III}] \otimes [\textrm{any}]$ with $\Lambda \ne 0$}
\newline
After substituting $\mu=0=\varkappa$, $\alpha_{0}=\textrm{const}\ne0$, $\nu=2 \alpha_{0}\Lambda$, $\Lambda \ne 0$ and $\gamma=\gamma(w,t)$ one brings the hyperheavenly equation to the form
\begin{equation}
\label{rownanie_zredukowane_natyp_III}
\frac{\Lambda}{6} F_{\phi \phi} + \tau F_{\phi t} + 3 \tau^{2} \alpha_{0}^{2} \phi^{5} + \gamma \phi =0
\end{equation}
with the general solution
\begin{equation}
F(\phi,w,t) = \mathfrak{f} \Big( \tau \phi- \frac{\Lambda}{6} \, t,w \Big) - \frac{3}{7 \Lambda} \tau^{2} \alpha_{0}^{2} \, \phi^{7} + \mathfrak{g}_{t} \, \phi^{2} - \frac{\Lambda}{3\tau} \, \mathfrak{g} \, \phi + \mathfrak{h} 
\end{equation}
where $\mathfrak{g}=\mathfrak{g}(w,t)$, $\mathfrak{h}=\mathfrak{h}(w,t)$ and $\mathfrak{f}$ are arbitrary functions of their variables. Moreover, $\gamma$ is given by $\gamma=-2 \tau \mathfrak{g}_{tt}$. The general solution for the key function $W$ reads
\begin{equation}
W(\eta,\phi,w,t) = \alpha_{0} \eta \phi^{3} + \mathfrak{f} \Big( \tau \phi- \frac{\Lambda}{6} \, t,w \Big) - \frac{3}{7 \Lambda} \tau^{2} \alpha_{0}^{2} \, \phi^{7} + \mathfrak{g}_{t} \, \phi^{2} - \frac{\Lambda}{3\tau} \, \mathfrak{g} \, \phi + \mathfrak{h} 
\end{equation}
The metric is
\begin{eqnarray}
\label{metryka_typu_III_any_IK3}
ds^{2} &=& (\phi \tau)^{-2} \bigg\{ 2\tau (d \eta \underset{s}{\otimes} d w - d \phi \underset{s}{\otimes} dt) +   \frac{\Lambda}{3}  \, dt \underset{s}{\otimes}dt  -8 \tau^{2} \alpha_{0} \phi^{3} \, dw \underset{s}{\otimes}dt
\\ \nonumber
&& \ \ \ \ \ \ \ \ \ \ 
  + 2 \Big(  -\tau^{4}  \mathfrak{f}_{zz} \, \phi+ 2 \tau^{3} \mathfrak{f}_{z} + \frac{12}{\Lambda} \tau^{4} \alpha_{0}^{2} \, \phi^{6} + 2 \tau^{2} \mathfrak{g}_{t} \, \phi - \frac{2}{3}\tau \Lambda \mathfrak{g}    \Big) dw \underset{s}{\otimes}dw \bigg\} \ \ \ \ \ \ 
\end{eqnarray}
where $z:= \tau \phi - \frac{\Lambda}{6} t$. 
\newline
Curvature
\begin{eqnarray}
\label{krzywizna_dla_typu_III_III}
&& C^{(3)}=0 \ \ \ , \ \ \ C^{(2)} = -2 \tau \alpha_{0} \Lambda \, \phi^{5} \ \ \ , \ \ \ C^{(1)}=-4 \tau^{2} \phi^{7} \big( \mathfrak{g}_{ttt} + \alpha_{0}^{2} \Lambda \, \phi^{3} \big)
\\ \nonumber
&& C_{\dot{A}\dot{B}\dot{C}\dot{D}} = \phi^{3} \, J_{(\dot{A}}J_{\dot{B}}J_{\dot{C}}L_{\dot{D})}
\end{eqnarray}
where 
\begin{equation}
\label{spinor_L_dla_typu_IIIxIII}
L_{\dot{A}} := \Big( \tau^{4} \mathfrak{f}_{zzzz} - \frac{360}{\Lambda} \tau^{2} \alpha_{0}^{2} \, \phi^{3} \Big) J_{\dot{A}} - 24 \alpha_{0} \, K_{\dot{A}}
\end{equation}
So the metric (\ref{metryka_typu_III_any_IK3}) describes the types $[\textrm{III}] \otimes [\textrm{III}]$ with $\Lambda \ne 0$.
\newline
($\alpha_{0}$ can be re-gauged to $1$ without any loss of generality).
\newline
\textbf{Type $[\textrm{N}] \otimes [\textrm{any}]$ with $\Lambda \ne 0$}
\newline
Here we have $\mu=\nu=\varkappa=\alpha=0$, $\gamma=\gamma(w,t)$, $\gamma_{t} \ne 0$, but cosmological constant is necessarily $\ne0$. This case can be easily obtained from the previous case by substituting $\alpha_{0}=0$ in formulas (\ref{rownanie_zredukowane_natyp_III}) - (\ref{spinor_L_dla_typu_IIIxIII}). The key function
\begin{equation}
W(\phi,w,t) =  \mathfrak{f} ( z,w)  + \mathfrak{g}_{t} \, \phi^{2} - \frac{\Lambda}{3\tau} \, \mathfrak{g} \, \phi + \mathfrak{h} 
\end{equation}
where $\mathfrak{g}=\mathfrak{g}(w,t)$, $\mathfrak{h}=\mathfrak{h}(w,t)$ and $\mathfrak{f}=\mathfrak{f}(z,w)$ are  arbitrary functions and $z:= \tau \phi - \frac{\Lambda}{6} t$. The nonzero curvature coefficients read
\begin{equation}
\label{krzywizna_dla_NN_z_Lambda}
C^{(1)}=-4 \tau^{2} \mathfrak{g}_{ttt} \, \phi^{7}   \ \ \ \ , \ \ \ \ 
C_{\dot{A}\dot{B}\dot{C}\dot{D}} = \tau^{4} \mathfrak{f}_{zzzz} \, \phi^{3} \, J_{\dot{A}}J_{\dot{B}}J_{\dot{C}}J_{\dot{D}}
\end{equation}
and the metric takes the following form
\begin{eqnarray}
\label{metryka_typu_N_N_IK3}
ds^{2} &=& (\phi \tau)^{-2} \bigg\{ 2\tau (d \eta \underset{s}{\otimes} d w - d \phi \underset{s}{\otimes} dt) +   \frac{\Lambda}{3}  \, dt \underset{s}{\otimes}dt 
\\ \nonumber
&& \ \ \ \ \ \ \ \ \ \ 
  + 2 \Big(  -\tau^{4}  \mathfrak{f}_{zz} \, \phi+ 2 \tau^{3} \mathfrak{f}_{z}  + 2 \tau^{2} \mathfrak{g}_{t} \, \phi - \frac{2}{3}\tau \Lambda \mathfrak{g}    \Big) dw \underset{s}{\otimes}dw \bigg\} \ \ \ \ \ \ 
\end{eqnarray}
with $\gamma=-2\tau \mathfrak{g}_{tt}$. If $\mathfrak{f}_{zzzz} \ne 0$ and $\mathfrak{g}_{ttt} \ne 0$ we have a complex metric (\ref{metryka_typu_N_N_IK3}) of the type $[\textrm{N}] \otimes [\textrm{N}]$ with $\Lambda \ne 0$.
\newline
It is worth-while to note, that the vector $\partial_{\eta}$ is a null Debever-Penrose vector. Indeed, the self-dual null string is defined by the Pfaff system $d q^{\dot{M}}=0$. The quadruple dotted Penrose spinor is proportional to $J_{\dot{A}}$ (compare (\ref{krzywizna_dla_NN_z_Lambda})), so the anti-self-dual null string is given by the Pfaff system $J_{\dot{B}} g^{A \dot{B}}=0$. Two sets of null strings intersect each other along the congruence of complex null geodesic. The tangent vector to this geodesic is tangent to both self-dual and anti-self-dual null strings. It is easy to check, that this tangent vector is proportional to $\frac{1}{\tau} J^{\dot{A}} \partial_{\dot{A}} = \partial_{\eta}$. An interesting symmetry arises: the Killing vector $\partial_{\eta}$ is a null multiple Debever - Penrose vector and it defines the congruence of the null complex geodesics. Unfortunatelly, the twist of this congruence vanishes. 
\newline
\textbf{Type $[\textrm{N}] \otimes [\textrm{any}]$ with $\Lambda=0$}
\newline
The structural functions read $\mu=\nu=\varkappa=0$, $\alpha=\alpha_{0}$, $\gamma=\gamma(w,t)$, $\gamma_{t} \ne 0$. The hyperheavenly equation takes the form
\begin{equation}
\label{rownaniehiperniebianskienatypnbezlambda}
\tau F_{\phi t} + 3 \tau^{2} \alpha_{0}^{2} \phi^{5} + \gamma \phi = 0
\end{equation}
One easily gets the general solution of (\ref{rownaniehiperniebianskienatypnbezlambda}) 
\begin{equation}
F(\phi,w,t) = - \frac{1}{2}\tau \alpha_{0}^{2} \phi^{6} t - \frac{1}{2\tau} \mathfrak{f} \, \phi^{2} + \mathfrak{g} + \mathfrak{h}
\end{equation}
where $\mathfrak{g}=\mathfrak{g}(\phi,w)$, $\mathfrak{h}=\mathfrak{h}(w,t)$, $\mathfrak{f}=\mathfrak{f}(w,t)$, and $\gamma=\mathfrak{f}_{t}$. Finally, the key function $W$ has the form
\begin{equation}
W(\eta,\phi,w,t) = \alpha_{0} \eta \phi^{3} - \frac{1}{2}\tau \alpha_{0}^{2} \phi^{6} t - \frac{1}{2\tau} \mathfrak{f} \, \phi^{2} + \mathfrak{g} + \mathfrak{h}
\end{equation}
The metric is
\begin{eqnarray}
\label{metryka_typu_N_III_bez_Lambdy}
ds^{2} &=& (\phi \tau)^{-2} \bigg\{ 2\tau (d \eta \underset{s}{\otimes} d w - d \phi \underset{s}{\otimes} dt)      -8 \tau^{2} \alpha_{0} \phi^{3} \, dw \underset{s}{\otimes}dt
\\ \nonumber
&& \ \ \ \ \ \ \ \ \ \ 
  + 2 \Big(  9 \tau^{3} \alpha_{0}^{2} \, t \phi^{5} - \tau \mathfrak{f} \, \phi - \tau^{2} \mathfrak{g}_{\phi\phi} \, \phi +2\tau^{2} \, \mathfrak{g}_{\phi}    \Big) dw \underset{s}{\otimes}dw \bigg\} \ \ \ \ \ \ 
\end{eqnarray}
and the curvature reads
\begin{eqnarray}
&&C^{(3)}=0 =C^{(2)}  \ \ \ , \ \ \ C^{(1)}=2 \tau \mathfrak{f}_{tt} \, \phi^{7}   
\\ \nonumber
&& C_{\dot{A}\dot{B}\dot{C}\dot{D}} =  \phi^{3} \, J_{(\dot{A}}J_{\dot{B}}J_{\dot{C}}L_{\dot{D})}
\end{eqnarray}
where
\begin{equation}
L_{\dot{A}} := \big( \mathfrak{g}_{\phi \phi \phi \phi} - 180\tau \alpha_{0}^{2} \, t \phi^{2} \big) J_{\dot{A}} - 24 \alpha_{0} \, K_{\dot{A}}
\end{equation}
So the allowed types are $[\textrm{N}] \otimes [\textrm{III,N}]$ and the type $[\textrm{N}] \otimes [\textrm{N}]$ appears when $\alpha_{0}=0$, $\mathfrak{f}_{tt} \ne 0$ and $\mathfrak{g}_{\phi \phi \phi \phi} \ne 0$. (In the case with nonzero $\alpha_{0}$, it can be gauged to 1).

\subsection{Examples of Osserman spaces with symmetry}

The hyperheavenly space theory is useful in analysis of, so called \textsl{Osserman spaces}. Considerations on the complex relativity remain valid for the case of the ultrahyperbolic relativity. The complex manifold reduces to the 4-dimensional real smooth Riemannian manifold with the metric $ds^{2}$ of signature $(++--)$. Instead of the complex holomorphic functions we deal with real smooth functions. 

Let $(x^{1}, x^{2}, x^{3}, x^{4})$ be local coordinates on some open neighborhood of $p \in \mathcal{M}$. If $R^{\alpha}_{\ \beta \gamma \delta}$ denote the components of the curvature tensor at $p$ with respect to the natural basis $\big( \frac{\partial}{\partial x^{\beta}} \big)_{p}$ $(\alpha, \beta, \gamma, \delta=1,...,4)$ and $X^{\beta}$ stand for the components of $X \in T_{p}\mathcal{M}$ in the same basis then one can define the following endomorphism
\begin{eqnarray}
&&\mathscr{R}_{(X)} : T_{p}\mathcal{M} \longrightarrow T_{p}\mathcal{M}
\\ \nonumber
&&(\mathscr{R}_{(X)} Y)^{\alpha} := R^{\alpha}_{\ \beta \gamma \delta} X^{\beta} X^{\gamma} Y^{\delta} \ \ , \ \ \ Y \in T_{p}\mathcal{M}
\end{eqnarray}
If $X \in S_{p}^{+} \mathcal{M}$, where $S_{p}^{+} \mathcal{M} := \{ X \in T_{p} \mathcal{M}: ds^{2} (X,X)=1 \} $, then this endomorphism is called the \textsl{Jacobi operator with respect to $X \in T_{p}\mathcal{M}$}.

$(\mathcal{M},ds^{2})$ is said to be \textsl{Osserman at $p \in \mathcal{M}$} if the characteristic polynomial of $\mathscr{R}_{(X)}$ is independent of $X \in S_{p}^{+} \mathcal{M}$. 
$(\mathcal{M},ds^{2})$ is called \textsl{pointwise Osserman} if $(\mathcal{M},ds^{2})$ is Osserman at each point $p \in \mathcal{M}$. The folowing theorem holds true: $(\mathcal{M},ds^{2})$ is pointwise Oserman if and only if $(\mathcal{M},ds^{2})$ is Einstein and self-dual (or anti-self-dual) \cite{biblio_60, biblio_33}. 

$(\mathcal{M},ds^{2})$ is called \textsl{globally Osserman} if the characteristic polynomial of $\mathscr{R}_{(X)}$ is independent of $X \in S^{+} (\mathcal{M})$ (where $S^{+} (\mathcal{M}) := \cup_{p \in \mathcal{M}} S_{p}^{+} \mathcal{M}$). The following theorem holds true: $(\mathcal{M},ds^{2})$ is globally Osserman iff for each point $p \in \mathcal{M}$ there exist an open neighbourhood $U$ of $p$ and an orientation on $U$ such that the conditions $C_{AB \dot{C}\dot{D}}=0=C_{\dot{A}\dot{B}\dot{C}\dot{D}}$ hold on $U$ and, moreover, the invariants $\stackrel{(2)}{\mathscr{C}} :=  C^{AB}_{\ \ \ CD} \, C^{CD}_{\ \ \ AB}$ and $\stackrel{(3)}{\mathscr{C}} := C^{AB}_{\ \ \ CD} \, C^{CD}_{\ \ \ EF} \, C^{EF}_{\ \ \ AB}$ are constant on $\mathcal{M}$ \cite{biblio_60, biblio_33}. It has been proved in \cite{biblio_33} that $\stackrel{(2)}{\mathscr{C}}$ and $\stackrel{(3)}{\mathscr{C}}$ are constant on $\mathcal{M}$ iff $\mu=0$, i.e., the corresponding hyperheavenly space is of the type $[\textrm{III, N}] \otimes [-]$.

Hyperheavenly spaces with $\Lambda$ and $C_{ABCD}=0$ (or $C_{\dot{A}\dot{B}\dot{C}\dot{D}}=0$) are pointwise Osserman.  If additionally $\Lambda \ne 0$ and anti-self-dual strings (self-dual strings, respectively) are expanding then the corresponding pointwise Osserman spaces are not Walker spaces. Self-duality (anti-self-duality) condition means, that pointwise Osserman spaces may admit conformal symmetries. We do not still have an algebraic tools advanced enough, to consider conformal symmetries. In order to deal with such symmetries, it is enough to take conformal factor in the form (\ref{conforemna_postac_czynnika_chi}) and $C_{ABCD}=0$. The analysis of the conformal symmetries in heavenly spaces with $\Lambda$ will be considered elsewhere. 

The main results presented here allow us to analyse only homothetic and isometric Killing symmetries in pointwise Osserman spaces. The first way is to take $C_{ABCD}=0$ (the general case, but still complicated key function $W$, being in general function of four variables). The second is to take $C_{\dot{A}\dot{B}\dot{C}\dot{D}}=0$. Of course, in that way we get only algebraically degenerate pointwise Osserman spaces, but the key function becomes the polynomial in $p^{\dot{M}}$. Both hyperheavenly equation and master equation can be separated into the system of the unknown functions of the two variables $q^{\dot{A}}$. Both ways will be considered in the work devoted to the symmetries in the Osserman spaces.

However, the simple example of the pointwise and, in fact, global (!) Osserman spaces admitting isometric Killing symmetry can be easily obtained from the metric (\ref{metryka_typu_N_N_IK3}). If $\mathfrak{f}$ is a third-order polynomial in $z$, $\mathfrak{f}_{zzzz} =0$, then the space becomes self-dual and globally Osserman. Moreover, if $\mathfrak{g}_{ttt} \ne 0$ the space is globally Osserman of the type $[\textrm{N}] \otimes [-]$ with $\Lambda \ne 0$. It is not a Walker space. 

The second example, with $\Lambda=0$ follows from the metric (\ref{metryka_typu_N_III_bez_Lambdy}). With $\alpha_{0}=0$, $\mathfrak{g}_{\phi \phi \phi \phi}=0$ and $\mathfrak{f}_{tt} \ne 0$ we deal with the pointwise Osserman space of the type $[\textrm{N}] \otimes [-]$. The self-dual null strings are expanding, but $\Lambda=0$ means, that there exists nonexpanding anti-self-dual null strings. The space is Walker on the anti-self-dual side.

\setlength\arraycolsep{2pt}
\setcounter{equation}{0}

\section{Detailed derivation of the master equation.}
\label{sekcja_szczegolowe_obliczenia}

In this section we present in some details the reduction of Killing equations to one master equation (\ref{expanding_master_equation}). Our basic assumption is, that the space is not conformally half-flat, i.e., $|C^{(3)}|+ |C^{(2)}|+ |C^{(1)}| \ne 0$. We already have shown that $\chi=\chi_{0}=\textrm{const}$, and 
\begin{equation}
\Lambda \chi_{0}=0
\end{equation}
Besides, from the integrability conditions $M_{ABCD}$ (\ref{wwarunki_MABCD}) it follows, that $\partial^{\dot{N}} (\phi^{2} k_{\dot{N}}) = 0$, from the Killing equations $E_{11}^{\ \ \dot{A}\dot{B}}$ (see (\ref{rowwnania_Killinga})), one finds that $\partial^{(\dot{A}} \big(\phi^{2} k^{\dot{B})} \big) = 0$. Consequently we have
\begin{equation}
k^{\dot{A}} = \phi^{-2} \, \delta^{\dot{A}}
\end{equation}
where $\delta^{\dot{A}}$ is an arbitrary function of $q^{\dot{M}}$ only.
\newline
From the definition of $l_{11}$ we obtain its explicit form (\ref{postac_l11}), which is consistent with $L_{111}^{\ \ \ \; \dot{A}}$. Second step is to investigate $L_{211}^{\ \ \ \; \dot{A}}$. Contracting it with $J_{\dot{A}}$ one can get very useful integrability condition (\ref{integrability_condition_2}). Contraction with $K_{\dot{A}}$ gives explicit form of $l_{12}$ (\ref{postac_l12}). With this form of $l_{12}$, the integrability conditions $L_{112}^{\ \ \ \; \dot{A}}$ become identities.

Knowing the form of $k_{\dot{A}}$, the second triplet of Killing equations $E_{12}^{\ \ \dot{A}\dot{B}}$ can be brought to the form
\begin{equation}
E_{12}^{\ \ \dot{A}\dot{B}} \equiv \partial^{(\dot{A}} V^{\dot{B})}=0 
\end{equation}
where
\begin{equation}
V^{\dot{A}} := h^{\dot{A}} - \delta_{\dot{S}} \, Q^{\dot{S}\dot{A}} + \frac{\partial \delta^{(\dot{S}}}{\partial q_{\dot{A})}} \, p_{\dot{S}}
\end{equation}
But if $\partial^{(\dot{A}} V^{\dot{B})}=0$ then $V^{\dot{A}} = V p^{\dot{A}} + \epsilon^{\dot{A}}$ with $V=V(q^{\dot{M}})$ and $\epsilon^{\dot{A}} = \epsilon^{\dot{A}} (q^{\dot{M}})$ being arbitrary functions of their variables. Hence
\begin{equation}
\label{postac_h}
h^{\dot{A}} - \delta_{\dot{S}} \, Q^{\dot{S}\dot{A}} =-  \frac{\partial \delta^{(\dot{S}}}{\partial q_{\dot{A})}} \, p_{\dot{S}} + V  p^{\dot{A}} + \epsilon^{\dot{A}}
\end{equation}

The final Killing equation, i.e., the $E$ equation can be rearranged to the form
\begin{equation}
\label{E_equation_in_different_form}
\frac{1}{2} E \equiv \phi^{4} \, \partial_{\dot{N}} \Big[ \phi^{-4} \big( h^{\dot{N}} - \delta_{\dot{S}} \, Q^{\dot{S}\dot{N}} \big) \Big] + 4 \chi_{0} - \frac{\partial \delta^{\dot{N}}}{\partial q^{\dot{N}}} = 0
\end{equation}
Inserting $h^{\dot{A}} - \delta_{\dot{S}} \, Q^{\dot{S}\dot{A}}$ from (\ref{postac_h}) into (\ref{E_equation_in_different_form}) one gets the polynomial in $\phi$, and, finally, the solutions for $V$ and $\epsilon^{\dot{A}}$
\begin{eqnarray}
&& 2V= 4\chi_{0} - \frac{\partial \delta^{\dot{N}}}{\partial q^{\dot{N}}} + \frac{4}{\tau} \, K_{\dot{S}}J_{\dot{N}} \, \frac{\partial \delta^{(\dot{S}}}{\partial q_{\dot{N})}}
\\ \nonumber
&&J_{\dot{N}}\epsilon^{\dot{N}} = 0 \ \ \ \rightarrow \ \ \ \epsilon^{\dot{N}} = \epsilon J^{\dot{N}}
\end{eqnarray} 
with $\epsilon = \epsilon(q^{\dot{M}})$ being an arbitrary function.
\newline
Putting this into (\ref{postac_h}) and bringing it to the most plausible form one obtains the final solution for $h^{\dot{A}}$
\begin{equation}
\label{solution_for_h}
h^{\dot{A}} - \delta_{\dot{S}} \, Q^{\dot{S}\dot{A}} = \bigg( 2\chi_{0} + \frac{2}{\tau} \, K_{\dot{S}}J_{\dot{N}} \, \frac{\partial \delta^{\dot{S}}}{\partial q_{\dot{N}}} \bigg) p^{\dot{A}} + \frac{\partial \delta_{\dot{S}}}{\partial q_{\dot{A}}} \, p^{\dot{S}} + \epsilon J^{\dot{A}}
\end{equation}
It is worth while to note that with above form of $h^{\dot{A}}$ the definition of $l_{12}$ is consistent with the form (\ref{postac_l12}). 

Equation $M_{1122}$ gives us the integrability condition (\ref{integrability_condition_3}), while the equation $M_{1222}$ gives (\ref{integrability_condition_4}). Of course, (\ref{integrability_condition_3}) is automatically satisfied for the types $\textrm{[III, N]}  \otimes \textrm{[any]}$ and it plays role only in the types $\textrm{[II, D]}  \otimes \textrm{[any]}$. Both (\ref{integrability_condition_3}) and (\ref{integrability_condition_4}) are automatically satisfied for the type $\textrm{[N]}  \otimes \textrm{[any]}$. 

The next step is to calculate the form of $l_{22}$ from $L_{212}^{\ \ \ \; \dot{A}}$. In $L_{212}^{\ \ \ \; \dot{A}}$ it is useful to replace the factor $\partial_{\dot{B}} l_{12}$ from $L_{112}^{\ \ \ \; \dot{A}}$ equation and do not calculate it directly from (\ref{postac_l12}). Except this, there appears the factor $Q^{\dot{A}\dot{B}}Q^{\dot{X}}_{\ \; \dot{B}}$ which is skew symmetric in the indices $\dot{A}\dot{X}$ and it can be changed according to
\begin{equation}
Q^{\dot{A}\dot{B}}Q^{\dot{X}}_{\ \; \dot{B}} = \frac{1}{2} \in^{\dot{A}\dot{X}}  Q^{\dot{S}\dot{B}}Q_{\dot{S}\dot{B}} =  \in^{\dot{A}\dot{X}} \phi^{2} \, \mathcal{T}
\end{equation}
Using the hyperheavenly equation in the form (\ref{expanding_hyperheavenly_equation_form1}) to replace $\mathcal{T}$ one can remove quadratic terms with the second derivative of the key function $W$. After some algebraic work, using (\ref{integrability_condition_3}) and contracting $L_{212}^{\ \ \ \; \dot{A}}$ with $K_{\dot{A}}$ one obtains $l_{22}$ in the form (\ref{postac_l22}) (contraction $J_{\dot{A}} \cdot L_{212}^{\ \ \ \; \dot{A}}$ is an identity).

The integrability condition $L_{122}^{\ \ \ \; \dot{A}}$ gives no additional information since it becomes an identity.

Eq. (\ref{postac_l22}) must be consistent with the definition of $l_{22}$ given by (\ref{definicje_spinorow_LAB}). It is not an identity. Denote
\begin{equation}
\label{Sigma}
\Sigma := -\pounds_{K} W - W \bigg( 4 \chi_{0} + \frac{4}{\tau} K_{(\dot{S}}J_{\dot{N})} \, \frac{\partial \delta^{\dot{S}}}{\partial q_{\dot{N}}} + \frac{1}{\tau} K_{\dot{N}}J_{\dot{S}} \, \frac{\partial \delta^{\dot{S}}}{\partial q_{\dot{N}}} - \frac{\partial \delta^{\dot{N}}}{\partial q^{\dot{N}}} \bigg)
\end{equation}
After some tedious calculations one shows, that the consistency condition takes the form
\begin{eqnarray}
\label{consistency_condition_between_l22}
J_{\dot{A}} \, \partial^{\dot{A}} \Sigma &=& -\frac{1}{2 \tau} \, \frac{\partial^{2} \delta^{\dot{S}}}{\partial q_{\dot{N}} \partial q_{\dot{R}}} \, (2K_{\dot{S}}J_{\dot{N}} + J_{\dot{S}}K_{\dot{N}}) p_{\dot{R}} 
\\ \nonumber
&&+ \frac{1}{2 \tau^{2}} \, \frac{\partial \delta^{\dot{S}}}{\partial q_{\dot{N}}} \, K_{\dot{N}}K_{\dot{S}} \, \bigg( \mu \phi^{3} - \frac{\Lambda}{3} \bigg) - \frac{1}{2} J_{\dot{A}} \frac{\partial \epsilon}{\partial q_{\dot{A}}}
\end{eqnarray}
Eq. (\ref{consistency_condition_between_l22}) plays a crucial role in the most important part of this work i.e. in an integration of the third triplet of the Killing equations $E_{22}^{\ \ \dot{A}\dot{B}}$. Putting the $h^{\dot{A}}$ given by (\ref{postac_h}) into $E_{22}^{\ \ \dot{A}\dot{B}}$, after some obvious cancellations one gets
\begin{equation}
\frac{1}{2} \phi^{-2} E_{22}^{\ \ \dot{A}\dot{B}} \equiv \frac{\partial R^{(\dot{A}}}{\partial q_{\dot{B})}} + Q^{\dot{S}(\dot{A}} \, \partial_{\dot{S}}R^{\dot{B})} - R^{\dot{S}} \, \partial_{\dot{S}} Q^{\dot{A}\dot{B}} + \frac{\partial \delta_{\dot{S}}}{\partial q_{(\dot{A}}} \, Q^{\dot{B})\dot{S}} + \delta^{\dot{S}} \, \frac{\partial Q^{\dot{A}\dot{B}}}{\partial q^{\dot{S}}} = 0
\end{equation}
where 
\begin{equation}
\label{definicja_era}
 R^{\dot{B}} :=\bigg( 2 \chi_{0} + \frac{2}{\tau} K_{\dot{S}} J_{\dot{N}} \frac{\partial \delta^{\dot{S}}}{\partial q_{\dot{N}}} \bigg) p^{\dot{B}} + \frac{\partial \delta_{\dot{S}}}{\partial q_{\dot{B}}} \, p^{\dot{S}} + \epsilon J^{\dot{B}}
\end{equation}
Taking $Q^{\dot{A}\dot{B}}$ in the form (\ref{expanding_QAB_z_Omega}) after some work we obtain
\begin{equation}
\frac{1}{2} \phi^{-2} E_{22}^{\ \ \dot{A}\dot{B}} \equiv \partial^{(\dot{A}} \Sigma^{\dot{B})} = 0
\end{equation}
where
\begin{eqnarray}
\label{Sigma_dotB}
\Sigma^{\dot{B}} &:=& -R^{\dot{S}} \, \partial_{\dot{S}} \Omega^{\dot{B}} + 2\Omega^{\dot{B}} \, \bigg( 2 \chi_{0} + \frac{2}{\tau} K_{\dot{S}} J_{\dot{N}} \frac{\partial \delta^{\dot{S}}}{\partial q_{\dot{N}}} \bigg)
+\frac{\partial \delta^{\dot{N}}}{\partial q^{\dot{N}}} \, \Omega^{\dot{B}} 
\\ \nonumber
&& + \frac{\partial \delta_{\dot{S}}}{\partial q_{\dot{B}}} \, \Omega^{\dot{S}} + \delta^{\dot{S}} \, \frac{\partial \Omega^{\dot{B}}}{\partial q^{\dot{S}}}
-\frac{\mu}{2 \tau^{3}} \phi^{4} \, \frac{\partial \delta^{\dot{S}}}{\partial q_{\dot{N}}} K_{\dot{S}}K_{\dot{N}} K^{\dot{B}} - \frac{\partial \epsilon}{\partial q_{\dot{B}}} \, \phi
\\ \nonumber
&& -\frac{\partial^{2} \delta^{\dot{S}}}{\partial q_{\dot{B}} \partial q_{\dot{R}}} \, p_{\dot{S}}p_{\dot{R}} + \frac{1}{2} \frac{\partial^{2} \delta^{\dot{B}}}{\partial q_{\dot{S}} \partial q_{\dot{R}}} \, p_{\dot{S}}p_{\dot{R}} + \frac{\partial}{\partial q^{\dot{T}}} \bigg( 2 \chi_{0} + \frac{2}{\tau} K_{\dot{S}} J_{\dot{N}} \frac{\partial \delta^{\dot{S}}}{\partial q_{\dot{N}}} \bigg)    p^{\dot{T}}p^{\dot{B}}
\end{eqnarray}
Inserting $\Omega^{\dot{B}}$ defined by (\ref{postac_Omegi}) into (\ref{Sigma_dotB}) one observes, that the terms containing the key function $W$ in (\ref{Sigma_dotB}) can be rearranged into the form $\phi^{4} \partial^{\dot{B}}(\phi^{-3}\Sigma)$ with $\Sigma$ given by (\ref{Sigma}). Moreover, after some algebraic tricks, using (\ref{integrability_condition_2}) one finds that
\begin{eqnarray}
&&-\frac{\partial^{2} \delta^{\dot{S}}}{\partial q_{\dot{B}} \partial q_{\dot{R}}} \, p_{\dot{S}}p_{\dot{R}} + \frac{1}{2} \frac{\partial^{2} \delta^{\dot{B}}}{\partial q_{\dot{S}} \partial q_{\dot{R}}} \, p_{\dot{S}}p_{\dot{R}} + \frac{\partial}{\partial q^{\dot{T}}} \bigg( 2 \chi_{0} + \frac{2}{\tau} K_{\dot{S}} J_{\dot{N}} \frac{\partial \delta^{\dot{S}}}{\partial q_{\dot{N}}} \bigg)    p^{\dot{T}}p^{\dot{B}}
=
\\ \nonumber
&&=\phi^{4} \, \partial^{\dot{B}} \bigg[ \phi^{-4} \frac{1}{2} \frac{\partial^{2} \delta^{\dot{S}}}{\partial q_{\dot{T}} \partial q_{\dot{R}}} \, p_{\dot{R}} p_{\dot{S}} p_{\dot{T}} \bigg]
\end{eqnarray}
Except this, from $\partial^{(\dot{A}} \Sigma^{\dot{B})} = 0$ it follows that $\Sigma^{\dot{B}} = \omega \, p^{\dot{B}} + \beta^{\dot{B}}$ (with $\omega$ and $\beta^{\dot{B}}$ being an arbitrary functions of the variable $q^{\dot{M}}$). Finally
\begin{eqnarray}
\label{Sigma_2}
\omega \, p^{\dot{B}} + \beta^{\dot{B}} &=& \phi^{4} \partial^{\dot{B}}(\phi^{-3}\Sigma) + \frac{\Lambda}{6 \tau^{2}} \bigg( \frac{1}{\tau} \frac{\partial \delta^{\dot{S}}}{\partial q_{\dot{N}}} K_{\dot{S}} K_{\dot{N}} \big( \eta J^{\dot{B}} - \phi K^{\dot{B}} \big) -\tau \epsilon \, K^{\dot{B}}   \bigg)
\\ \nonumber
&&-\frac{\mu}{2 \tau^{3}} \phi^{4} \, \frac{\partial \delta^{\dot{S}}}{\partial q_{\dot{N}}} K_{\dot{S}}K_{\dot{N}} K^{\dot{B}} - \frac{\partial \epsilon}{\partial q_{\dot{B}}} \, \phi
+\phi^{4} \, \partial^{\dot{B}} \bigg[ \phi^{-4} \frac{1}{2} \frac{\partial^{2} \delta^{\dot{S}}}{\partial q_{\dot{T}} \partial q_{\dot{R}}} \, p_{\dot{R}} p_{\dot{S}} p_{\dot{T}} \bigg]
\end{eqnarray}
Contracting (\ref{Sigma_2}) with $J_{\dot{B}}$ and using (\ref{consistency_condition_between_l22}) we get
\begin{eqnarray}
\label{solutions_for_omega_and_beta}
\omega &=& - \frac{3}{2} J^{\dot{A}} \frac{\partial \epsilon}{\partial q^{\dot{A}}} - \frac{\Lambda}{3 \tau^{2}}
 \frac{\partial \delta^{\dot{S}}}{\partial q_{\dot{N}}} K_{\dot{S}}K_{\dot{N}}
\\ \nonumber
\beta^{\dot{B}} &=& - \frac{\Lambda \epsilon}{6 \tau} K^{\dot{B}} - 3 \beta \, J^{\dot{B}}
\end{eqnarray}
(where $\beta=\beta(q^{\dot{M}})$ is an arbitrary function).
\newline
(An alternative way to obtain (\ref{solutions_for_omega_and_beta}) is to multiply (\ref{Sigma_2}) by $\phi^{-4}$ and derive $\partial_{\dot{B}} (\ref{Sigma_2})$)
\newline
Using the solutions for $\omega$ and $\beta^{\dot{B}}$ in Eq. (\ref{Sigma_2}) one can bring it to the form
\begin{eqnarray}
0&=&\phi^{4} \partial^{\dot{B}} \bigg\{ \phi^{-3} \bigg[ \Sigma + \beta + \frac{1}{2} \, p^{\dot{A}} \frac{\partial \epsilon}{\partial q^{\dot{A}}} - \frac{1}{2 \tau^{3}} \frac{\partial \delta^{\dot{S}}}{\partial q_{\dot{N}}} K_{\dot{S}}K_{\dot{N}} \, \eta \bigg( \mu \phi^{3} - \frac{\Lambda}{3} \bigg) 
\\ \nonumber
&& \ \ \ \ \ \ \ \ \ \ \ \ \ \ \ \ \ 
+ \frac{1}{2} \frac{\partial^{2} \delta^{\dot{S}}}{\partial q_{\dot{T}} \partial q_{\dot{R}}} \, p_{\dot{R}}p_{\dot{S}}p_{\dot{T}} \, \phi^{-1} \bigg] \bigg\}
\end{eqnarray}
Hence
\begin{eqnarray}
&&  \Sigma + \beta + \frac{1}{2} \, p^{\dot{A}} \frac{\partial \epsilon}{\partial q^{\dot{A}}} - \frac{1}{2 \tau^{3}} \frac{\partial \delta^{\dot{S}}}{\partial q_{\dot{N}}} K_{\dot{S}}K_{\dot{N}} \, \eta \bigg( \mu \phi^{3} - \frac{\Lambda}{3} \bigg) 
\\ \nonumber
&&\ \ \ \ \ \ + \frac{1}{2} \frac{\partial^{2} \delta^{\dot{S}}}{\partial q_{\dot{T}} \partial q_{\dot{R}}} \, p_{\dot{R}}p_{\dot{S}}p_{\dot{T}} \, \phi^{-1} =-  \alpha \, \phi^{3}
\end{eqnarray}
(where $\alpha=\alpha(q^{\dot{M}})$ is an arbitrary function).
\newline
Finally, using (\ref{Sigma}) we get the master equation.
\newline
In \cite{biblio_52} the authors obtained final result in a slightly different form, using a different potentialization of the $Q^{\dot{A}\dot{B}}$
\begin{equation}
\nonumber
Q^{\dot{A}\dot{B}} = \phi^{3} \, \partial^{(\dot{A}} A^{\dot{B})} 
+ \frac{\Lambda}{6\tau^{2}} K^{\dot{A}} K^{\dot{B}}
\end{equation}
where 
\begin{equation}
A^{\dot{A}} = - \phi^{-2} \partial^{\dot{A}}W + \frac{\mu}{\tau^{2}} K^{\dot{A}} \eta
\end{equation}
However, that way is leading to the same result (\ref{expanding_master_equation}) so it seems that there is a misprint in master equation in \cite{biblio_52}.
\newline
Integrability conditions $M_{2222}$ and $L_{222}^{\ \ \ \; \dot{A}}$ offer crazy calculations. It seems, that they contain the key function $W$, but the function $W$ can be removed from these conditions, using master equation (in $M_{2222}$) and both master and hyperheavenly equations (in $L_{222}^{\ \ \ \; \dot{A}}$), so they become polynomials in $p^{\dot{A}}$. After cancellations we find that $K_{\dot{A}} L_{222}^{\ \ \ \; \dot{A}}$ gives  (\ref{integrability_condition_5}), $J_{\dot{A}} L_{222}^{\ \ \ \; \dot{A}}$ is an identity and $M_{2222}$  gives (\ref{integrability_condition_6}) and (\ref{integrability_condition_7}). Note, that (\ref{integrability_condition_7}) involves $\Lambda$ and has been not presented in \cite{biblio_52}. Originally we have obtained (\ref{integrability_condition_7}) in the form
\begin{eqnarray}
\nonumber
&& J^{\dot{N}}  \frac{\partial}{\partial q^{\dot{N}}} \bigg( \frac{\tau}{2} \epsilon \nu + 3 \mu \beta \bigg) + \frac{\partial}{\partial q^{\dot{N}}} \bigg( J^{\dot{N}} \delta^{\dot{B}} \frac{\partial \gamma}{\partial q^{\dot{B}}} + 2 J_{\dot{B}} \delta^{\dot{B}} \frac{\partial \gamma}{\partial q_{\dot{N}}} \bigg)
 \ \ \ \ \ \ \ \ \ \ \ \ \ \ \ 
\\ \nonumber
&&  \ \ \ \ \ \ \ \ \ \ \ \ \ \ \ 
-\frac{\Lambda}{6 \tau} \bigg[ \tau \delta^{\dot{N}} \frac{\partial \varkappa}{\partial q^{\dot{N}}} - \epsilon K^{\dot{N}} \frac{\partial \mu}{\partial q^{\dot{N}}} + \nu \, K_{\dot{N}}K_{\dot{S}} \frac{\partial \delta^{\dot{S}}}{\partial q_{\dot{N}}} + \tau \varkappa \bigg( 2\, \frac{\partial \delta^{\dot{N}}}{\partial q^{\dot{N}}} + \frac{1}{\tau} J_{\dot{N}}K_{\dot{S}} \frac{\partial \delta^{\dot{S}}}{\partial q_{\dot{N}}} \bigg) \bigg]
\end{eqnarray}
but with help of (\ref{integrability_condition_5}) it can be brought to the form (\ref{integrability_condition_7}).

\section{Concluding remarks}

In this paper the problem of Killing symmetries in expanding hyperheavenly spaces with $\Lambda$ has been considered. We have found the explicit form of the Killing equations and their integrability conditions. The results of \cite{biblio_52} have been generalized to the case of nonzero cosmological constant and some small misprints in \cite{biblio_52} have been corrected. Finally, some ways of simplifying the hyperheavenly equation have been presented and interesting complex metrics have been calculated.

Metrics calculated in subsection \ref{subsekcja_na_typy} admit the Killing vector $\partial_{\eta}$. In that case all the nonlinearities in the hyperheavenly equation vanish and the explicite solution could be easily found. Especially interesting is the metric (\ref{metryka_typu_N_N_IK3}). It describes the type $[\textrm{N}] \otimes [\textrm{N}]$ with nonzero cosmological constant $\Lambda$. The Killing vector is null and it is the Debever - Penrose vector. The twist of the null, complex congruence defined by the Killing vector $\partial_{\eta}$ vanishes. Are there any possibilities to find the solutions of the complex type $[\textrm{N}] \otimes [\textrm{N}]$ admitting other Killing vectors, namely $\partial_{w}$ or $\partial_{t}$? We are going to answer these questions soon. 

The Killing symmetries in hyperheavenly spaces left some other, unsolved problems.

The classification of the Killing vectors in nonexpanding hyperheavenly spaces was not presented in \cite{biblio_51}. In both papers describing the Killing symmetries in nonexpanding hyperheavenly spaces (\cite{biblio_51}, \cite{biblio_27}) the ways of simplifying the final formulas has been presented, but the main criterion taken into account for classification has been formulated on the algebraic type of the Weyl tensor. It seems, that the more efficient way is to deal with the properties of the spinors $l_{AB}$ as a main criterion. That way was chosen in classification of the heavenly spaces made by Plebański and Finley in \cite{biblio_16} and also by us in the present paper. Are there any chances to obtain the similar classification in nonexpanding hyperheavenly spaces with $\Lambda$? One of the main result of \cite{biblio_16} is the new physical equation, called the Boyer-Plabański-Finley equation. It is the nonlinear differential equation for one function depending on three variables. It seems that there should exist some nonexpanding hyperheavenly equivalent of this equation. Moreover, the nonexpanding hyperheavenly spaces plays an important role in analysis of Einstein - Walker spaces. The symmetry of such spaces could be easily obtained from the Killing symmetry in nonexpanding hyperheavenly spaces.

The next important task is to find the master equation for the heavenly spaces with cosmological constant. Nonzero cosmological constant implies, that if $C_{ABCD}=C_{AB \dot{C}\dot{D}}=0$, then there does not exist any nonexpanding congruence of the self-dual null strings. The case of isometric Killing symmetries in such spaces was considered in \cite{biblio_53, biblio_54} but homothetic and conformal symmetries of that spaces are nowadays, unknown. Can the Killing equations be reduced to one master equation? Results of \cite{biblio_51} generalize the previous work of Plebański and Finley \cite{biblio_15} and prove that the conformal symmetries in heavenly spaces generate some new function of all variables, together with the first integral of the heavenly equation. Does the similar function and first integral appear in the master equation in heavenly spaces with cosmological constant? The analysis of the master equation for the heavenly spaces with $\Lambda$ allows to find various symmetries of the Osserman spaces.

The problems of conformal, homothetic and isometric Killing symmetries in heavenly spaces with cosmological constant, classification of the nonexpanding hyperheavenly spaces admitting Killing vector and symmetries in Osserman and Walker spaces will be considered in the next parts of our work.
\newline
\newline
\textbf{Acknowledgments.} 
\newline
The author is indebted to prof. M. Przanowski for many enlightning discussions and help in many crucial matters.
\newline
\newline
\textbf{Correctum.} 
There was found a few misprints in \cite{biblio_51}.
\newline
Formula (2.6)
\begin{eqnarray*}
\textrm{there is} \ \ && \ \ 
ds^{2} = \phi^{-2} \, (- dp^{\dot{A}} \underset{s}{\otimes} dq_{\dot{A}} + 
           Q^{\dot{A} \dot{B}} \, dq_{\dot{A}} \underset{s}{\otimes} dq_{\dot{B}})
\\ \nonumber
\textrm{there should be} \ \ && \ \ 
ds^{2} =2 \phi^{-2} \, (- dp^{\dot{A}} \underset{s}{\otimes} dq_{\dot{A}} + 
           Q^{\dot{A} \dot{B}} \, dq_{\dot{A}} \underset{s}{\otimes} dq_{\dot{B}})
\end{eqnarray*}
Formula (2.27)
\begin{eqnarray*}
\textrm{there is} \ \ && \ \ 
N^{\dot{A} \dot{B}} := D_{\dot{X}}^{\ \ \dot{A}} D_{\dot{Y}}^{\ \ \dot{B}} 
\bigg(  
\frac{2}{3} \, F'^{( \dot{X}} \sigma^{\dot{Y})} 
- \frac{\partial \sigma^{(\dot{X}}}{\partial q'_{\dot{Y})}}
+ \frac{1}{3} \, \Lambda \, \sigma^{\dot{A}} \sigma^{\dot{B}}
 \bigg)
\\ \nonumber
\textrm{there should be} \ \ && \ \
N^{\dot{A} \dot{B}} := D_{\dot{X}}^{\ \ \dot{A}} D_{\dot{Y}}^{\ \ \dot{B}} 
\bigg(  
\frac{2}{3} \, F'^{( \dot{X}} \sigma^{\dot{Y})} 
- \frac{\partial \sigma^{(\dot{X}}}{\partial q'_{\dot{Y})}}
+ \frac{1}{3} \, \Lambda \, \sigma^{\dot{X}} \sigma^{\dot{Y}}
 \bigg)
\end{eqnarray*}
Formula (4.13)
\begin{eqnarray*}
\textrm{there is} \ \ && \ \ 
l_{22} = - \frac{\partial}{\partial q^{\dot{M}}} ( 2\chi + b) + 2 \delta_{\dot{M}}N^{\dot{M}} - \frac{\partial \epsilon^{\dot{M}}}{\partial q^{\dot{M}}}
\\ \nonumber
\textrm{there should be} \ \ && \ \ l_{22} = - \frac{\partial}{\partial q^{\dot{M}}} ( 2\chi + b) \, p^{\dot{M}} + 2 \delta_{\dot{M}}N^{\dot{M}} - \frac{\partial \epsilon^{\dot{M}}}{\partial q^{\dot{M}}}
\end{eqnarray*}
Formula (4.15)
\begin{eqnarray*}
\textrm{there is} \ \ && \ \ 
K^{\dot{B}} = \delta^{\dot{M}} \Theta_{p^{\dot{M}}p_{\dot{B}}} - \frac{\partial \delta^{\dot{M}}}{\partial q_{\dot{B}}} \, p_{\dot{M}} + \frac{2}{3} \delta_{\dot{M}} F^{(\dot{M}} p^{\dot{B})} - \epsilon^{\dot{B}}
\\ \nonumber
\textrm{there should be} \ \ && \ \
K^{\dot{B}} = \delta^{\dot{M}} \Theta_{p^{\dot{M}}p_{\dot{B}}} - \frac{\partial \delta^{\dot{M}}}{\partial q_{\dot{B}}} \, p_{\dot{M}} + \frac{2}{3} \delta_{\dot{M}} F^{(\dot{M}} p^{\dot{B})} - \epsilon^{\dot{B}} - 2 \chi \, p^{\dot{B}}
\end{eqnarray*}
The first line under the formula (4.29)
\begin{eqnarray*}
\textrm{there is} \ \ && \ \ q_{\dot{A}} \frac{\partial a}{\partial q_{\dot{A}}} - 2 f_{1} =0
\\ \nonumber
\textrm{there should be} \ \ && \ \ q_{\dot{A}} \frac{\partial f_{1}}{\partial q_{\dot{A}}} - 2 f_{1} =0
\end{eqnarray*}

\end{document}